\documentclass[superscriptaddress,aps,preprintnumbers,amsmath,amssymb,prd,nofootinbib,preprint,longbibliography]{revtex4-1}
\pdfoutput=1

\usepackage{cancel}
\usepackage{graphicx,array}
\usepackage{color}
\usepackage{latexsym}
\usepackage{amsthm}
\usepackage{amsmath}
\usepackage{amssymb}
\usepackage[colorlinks=true,citecolor=blue,linkcolor=red]{hyperref}
\usepackage{bbold}
\usepackage{mathtools}
\usepackage{enumitem}
\usepackage{makecell}
\usepackage[table]{xcolor}
\usepackage{multirow}
\usepackage{booktabs}
\usepackage{setspace}
\usepackage{stackengine}
\usepackage{comment}
\usepackage{tikz}
\usepackage{verbatim}
\usepackage[normalem]{ulem}
\usepackage{cancel}
\usepackage{tabularx}

\numberwithin{equation}{section}

\makeatletter
\renewcommand{\p@subsection}{}
\renewcommand{\p@subsubsection}{}
\makeatother

\def\simgt{\mathrel{\lower2.5pt\vbox{\lineskip=0pt\baselineskip=0pt
           \hbox{$>$}\hbox{$\sim$}}}}
\def\simlt{\mathrel{\lower2.5pt\vbox{\lineskip=0pt\baselineskip=0pt
           \hbox{$<$}\hbox{$\sim$}}}}

\newcommand{\be}{\begin{equation}}
\newcommand{\ee}{\end{equation}}
\newcommand{\bea}{\begin{eqnarray}}
\newcommand{\eea}{\end{eqnarray}}

\newcommand{\gsim}{\lower.7ex\hbox{$\;\stackrel{\textstyle>}{\sim}\;$}}
\newcommand{\lsim}{\lower.7ex\hbox{$\;\stackrel{\textstyle<}{\sim}\;$}}

\definecolor{nicered}{rgb}{0.7,0.1,0.1}
\definecolor{nicegreen}{rgb}{0.1,0.5,0.1}
\definecolor{PatColor}{rgb}{0,.8,0}
\hypersetup{colorlinks,citecolor= blue,linkcolor= black}

\definecolor{violet}{rgb}{0.5,0.,1.0}

\begin{document}

\title{Warm Dark Matter meets Cold Dark Matter Isocurvature}

\author{Sai Chaitanya Tadepalli}
\thanks{For correspondence: saictade@iu.edu}
\affiliation{Physics Department, Indiana University, Bloomington, IN, 47405, USA}
\author{Tomo Takahashi}
\thanks{}
\affiliation{Department of Physics, Saga University, Saga 840-8502, Japan \vspace{1cm} }

\begin{abstract}
Isocurvature fluctuations can be generated in various scenarios in the early Universe. In particular, some specific models predict those with a blue-tilted spectrum, which is consistent with the constraints from cosmic microwave background such as Planck, although isocurvature fluctuations with an almost scale-invariant spectrum are severely constrained. We argue that cold dark matter (CDM) isocurvature fluctuations with blue-tilted spectrum are not only consistent with current cosmological data, but also can loosen the bound on the masses of warm dark matter (WDM), which suppresses small-scale power. In pure thermal WDM models with the adiabatic initial condition,  a combination of the data from Lyman-$\alpha$, gravitational lensing, and Milky Way satellites gives a lower bound on the WDM mass as $6~{\rm keV}$ at $95\%$ C.L. while mixed WDM+CDM models loosen these bounds to $m_{\rm WDM}\sim1$ keV for a warm-fraction $f_{\rm WDM}\lesssim0.14$ and $m_{\rm WDM}\sim600$ eV for $f_{\rm WDM}\lesssim0.08$. On the other hand, as we demonstrate, WDM scenarios with a blue-tilted CDM isocurvature power spectrum, even with only $1\%$ CDM contribution ($f_{\rm WDM}\sim0.99$), can allow WDM masses as low as $600$ eV. We further assess the implications of this ``warm + cold-isocurvature'' extension for the small-scale structure by performing $N$-body simulations, particularly focusing on nonlinear matter power spectrum and halo mass function.

\end{abstract}

\maketitle

\vspace{1cm}

\begingroup
\hypersetup{linkcolor=black}
\renewcommand{\baselinestretch}{1.26}\normalsize
\tableofcontents
\renewcommand{\baselinestretch}{2}\normalsize
\endgroup

\newpage

\section{Introduction}

The standard cosmological model, $\Lambda$CDM, successfully describes a wide array of large-scale structure (LSS) observations and cosmic microwave background (CMB) anisotropies~\cite{Planck:2019nip}. However, persistent discrepancies remain, such as in the density profile observed in the innermost regions of galaxies \cite{deBlok:2009sp}, and observational tensions on small and intermediate scales probed by Lyman-$\alpha$ forest data \cite{Rogers:2023upm} and dwarf galaxy abundances \cite{Klypin:1999uc,Moore:1999nt,Simon:2019nxf,Sales:2022ich}. See Refs.~\cite{DelPopolo:2016emo,Weinberg:2013aya} for a review of small-scale problems in $\Lambda$CDM and Ref.~\cite{CosmoVerse:2025txj} for existing cosmological tensions. These challenges have supported investigations into alternative dark matter (DM) models and non-standard initial conditions. Warm dark matter (WDM) scenarios, characterized by a finite particle velocity dispersion, naturally suppress structure formation below a characteristic free-streaming scale $\lambda_{\rm FS}$, potentially alleviating small-scale tensions inherent to pure cold dark matter (CDM) models. Such particles can be realized as axinos  \cite{Covi:2009pq} (for thermally produced  axinos, see Ref.~\cite{Strumia:2010aa}), gravitinos \cite{FAYET1977461}, and sterile neutrinos \cite{Dodelson:1993je,Shi:1998km} (see Ref.~\cite{Jaramillo:2022mos} for thermal production). At scales larger than $\lambda_{\rm FS}$, WDM closely resembles CDM in its structure formation morphology. For instance, WDM particles with masses of $\sim \mathcal{O}(\rm keV)$ do not cluster efficiently on or below dwarf galaxy scales. Constraints on pure WDM models have been extensively explored across various observational probes with similar lower limits, including the recent high-resolution Lyman-$\alpha$ measurements  \cite{Irsic:2023equ}, Milky Way satellite counts  \cite{Nadler:2021dft,Tan:2024cek}, luminosity functions of distant galaxies \cite{Menci:2016eui,Rudakovskyi:2021jyf}, strong gravitational lensing \cite{Gilman:2019nap} to name a few. A combined analysis incorporating Lyman-$\alpha$, strong gravitational lensing and Milky Way satellites excludes thermal WDM relics lighter than $6.0$ keV at  2$\sigma$  confidence level \cite{Enzi:2020ieg}.

These stringent constraints on pure WDM models primarily arise from the significant suppression of small-scale power due to the free-streaming of WDM particles. With recent studies pushing the lower limits on the WDM particle mass even higher, it raises doubts about the continued viability of pure WDM scenarios as a solution to the small-scale challenges of CDM structure formation. For instance, in Ref.~\cite{Schneider:2013wwa} the author questioned whether WDM could still effectively address the \textit{too-big-to-fail} (TBTF) problem \cite{Boylan-Kolchin:2011qkt}.
Similarly, Ref.~\cite{Maccio:2012qf}  showed that for WDM masses $1-2$~keV, the expected core is $\mathcal{O}(10-20)$~pc for DM halos, far lower than the expected $\sim 1$ kpc. A minimal extension of WDM is to assume that DM consists of both cold and warm components. Mixed WDM+CDM scenarios relax the lower bounds on WDM particle masses while simultaneously shifting the suppression scale to smaller wavenumbers (i.e., larger spatial scales), but with a considerably shallower suppression of the matter power spectrum. This allows for the possibility of retaining the beneficial suppression effects of WDM on small-scale structure formation while simultaneously ensuring sufficient clustering power through a CDM component to match observational data. Such hybrid models could thus reconcile the strengths of both dark matter types, addressing the persistent discrepancies observed on small scales without violating constraints from current cosmological observations. This has spurred growing interest, with several recent studies exploring their implications: Ref.~\cite{Parimbelli:2021mtp} demonstrated notable effects on weak lensing, galaxy clustering, and halo mass functions; Ref.~\cite{Garcia-Gallego:2025kiw} utilized Lyman-$\alpha$ forest measurements at high redshift to constrain these models; Ref.~\cite{Tan:2024cek} examined their viability through Milky Way satellite galaxy counts; Ref.~\cite{Schneider:2018xba} gave constrain using global 21-cm signal; and Ref.~\cite{Kamada:2016vsc} investigated constraints from anomalous strong lensing systems.
 
For sub‑keV thermal WDM, the allowed warm fraction must be low to satisfy observational constraints. To illustrate current limits, a thermal WDM particle with a mass of approximately $1$~keV is constrained to contribute  $\lesssim 14\%$ of the total dark matter density in mixed WDM+CDM models. While mixed WDM+CDM models mitigate some of the tensions by altering the composition of dark matter, the allowed low warm fraction for sub-keV WDM yield halos that remain centrally cuspy (with only modestly outer cores) and produce shallower small-scale suppression than pure WDM. Consequently, lowering the WDM mass further to attempt to produce larger cores tends to conflict with observations unless accompanied by strong baryonic feedback \cite{Anderhalden:2012qt,Lovell:2013ola,Parimbelli:2021mtp}.
This raises the question of whether the allowed WDM fraction can be increased through a complementary mechanism. In this context, modifying the primordial perturbation spectrum offers a promising avenue. In particular, isocurvature perturbations in the CDM component (CDI) provide a mechanism to adjust small-scale clustering, decoupled from the particle-velocity-based suppression in WDM. While observational constraints from the CMB data \cite{Planck:2018jri,ACT:2025tim} stringently limit the amplitude of isocurvature modes on large scales, they leave open the intriguing possibility of substantial small-scale isocurvature power, if the initial CDI spectrum is strongly blue-tilted.

This motivates us to explore cosmological scenarios where both effects coexist, forming a mixed WCDM+CDI model. In this framework, thermal WDM suppresses power on small scales due to free-streaming, while a blue-tilted uncorrelated CDI spectrum can inject compensatory power precisely on these scales. By appropriately tuning the amplitude and spectral index of CDI fluctuations, one can partially or fully compensate for the suppressed power on relevant scales, potentially permitting a larger WDM fraction for lower WDM particle masses without conflicting with observational data. In short, a percent-level CDI admixture can plausibly mask the small-scale power deficit while keeping some WDM-like concentration relief. One could also consider a WDM+BI cosmology consisting of pure WDM dark matter and baryonic blue-tilted isocurvature (BI). The presence of baryonic fluid leads to added complications in $N$-body simulations, and their inherent gas pressure at precisely the small scales of interest. We therefore restrict our analysis to the WCDM+CDI framework.

Admittedly, introducing additional parameters such as the CDI amplitude and tilt naturally increases model complexity, which should be weighed carefully against Occam’s razor via a full Bayesian analysis. However, this extension is physically well-motivated, since known CDM candidates such as QCD axions or axion-like particles (ALPs) can intrinsically source large blue-tilted isocurvature perturbations \cite{Kasuya:2009up,Chung:2024ctx}. Exploring the interplay between WDM and CDI thus provides a targeted probe into early-Universe physics, while offering avenues for rigorously testing whether primordial perturbations can alleviate persistent tensions in small-scale structure formation.

A comprehensive investigation of WCDM+CDI models would necessitate sophisticated hydrodynamic $N$-body simulations and extensive marginalization over all new parameters, exceeding the scope of this preliminary study. Instead, we adopt a simpler approach by performing DM-only $N$-body simulations, focusing on the feasibility of a CDI component, significantly relaxing current WDM constraints from probes sensitive to the underlying small-scale nonlinear matter power spectrum and halo statistics.  

The order of the presentation is as follows: in Sec.~\ref{sec:Theory_setup} we give a brief overview of primordial initial conditions focusing on CDM isocurvature fluctuations, followed by a discussion of the mixed warm and cold DM cosmological model. Therein, we provide an analytic approximation for the linear matter transfer function of the mixed DM-model in Sec.~\ref{Sec:WCDM}. In Sec.~\ref{sec:WCDM+CDI}, we discuss the mixed WCDM+CDI model using linear perturbation theory in LSS. We present a joint CMB and BAO based constraint on the WCDM+CDI model in Sec.~\ref{sec:cmb_bao}, followed by the results on the compensation within the linear matter power from the additional CDI component in Sec.~\ref{sec:WCDM+CDI:linear_mPK}. In Sec.~\ref{sec:WCDM_NL}, we discuss results from the $N$-body simulation for a few fiducial mixed WCDM+CDI models, presenting the nonlinear transfer function in Sec.~\ref{sec:Tk_NL} and halo mass function in Sec.~\ref{sec:hmf}. Finally, we conclude with a discussion in Sec.~\ref{sec:conclusion}.

\section{Theoretical Setup}\label{sec:Theory_setup}
Throughout this section, we work in the linear perturbation regime, describing fluctuations in the total matter field $\delta_m$. We begin with an overview of primordial initial conditions in Sec~\ref{sec:CDMiso} focusing on CDM isocurvature fluctuations, followed by the mixed warm and cold DM cosmological model in Sec.~\ref{Sec:WCDM}. We restrict our analysis to fermionic WDM particles that were in thermal equilibrium with the hot, dense primordial plasma. Consequently, their momentum distribution is Fermi-Dirac, and their background properties (free-streaming scale, velocity dispersion, relic abundance) can be solely determined from their masses. Upon decoupling, they free-stream, becoming non-relativistic once their temperature drops below their mass. For a $\mathcal{O}(100)$eV WDM, this non-relativistic transition occurs at $z\sim10^7$.

\subsection{Isocurvature initial conditions}\label{sec:CDMiso}
In studying the evolution of cosmological perturbations, one must solve a coupled system of linearized Einstein and fluid equations subject to appropriate initial conditions
\cite{Ma:1995ey,dodelson2020modern}. The initial conditions for the scalar fluctuations are divided into two classes: \textit{adiabatic} or \textit{isocurvature}
~\cite{Mukhanov:1990me,Bucher:1999re}. 
During inflation~(\cite{Guth:1980zm,Starobinsky:1980te,Sato:1980yn}), quantum fluctuations in the inflaton field generate curvature perturbations $\zeta$ which, via gravity (metric), source energy density fluctuations $\delta \rho$ in all other particle species. Meanwhile, light fields present during inflation can also acquire quantum fluctuations, which give rise to a variation in their local number-to-entropy ratio, $\delta(n/s)$ where $s$ is the entropy
density.

More precisely, adiabatic (isentropic) perturbations are fluctuations in
energy density and not particle number per unit entropy so that $\delta (n_i/s) = 0$ for any species $i$. In contrast, isocurvature (entropic) perturbations satisfy $\delta (n_i/s) \neq 0$, indicating relative fluctuations in the number density of species $i$ at fixed total energy density. For non-relativistic matter-like species $m$, one can show 
\begin{equation}
    \delta \left(\frac{n_m}{s}\right)=\delta_m - \frac{3}{4}\delta_r
\end{equation}where $\delta_{m(r)}$ corresponds to fluctuations in the matter (radiation) energy densities. In terms of the conserved superhorizon curvature perturbation $\zeta$, the gauge-invariant isocurvature fluctuation between two species $i$ and $j$ is given in Ref.~\cite{Langlois:2008vk} as
\begin{equation}
    \mathcal{S}_{ij} = 3 \left(\zeta_i - \zeta_j\right),
\end{equation}while for the superhorizon adiabatic fluctuation, it follows that
\begin{equation}
    \zeta_i = \zeta_j.
\end{equation} 
By convention, one measures isocurvature fluctuations with respect to radiation, hence CDM isocurvature is expressed as \[\mathcal{S}_{\rm c}= 3 \left(\zeta_{\rm c} - \zeta_r\right).\]

The primordial (conserved superhorizon) dimensionless power spectrum $\Delta^2(k)$ of the adiabatic and isocurvature fluctuations can be parametrized in terms of a standard power-law expression as
\begin{eqnarray}
    \Delta^2_\mathcal{R}(k)&=\displaystyle\frac{k^3}{2\pi^2}P_\mathcal{R}(k)&=
    A_{\rm ad}(k_p)
    \left(\frac{k}{k_p}\right)^{n_{\rm ad}-1},\label{Eq:PR}\\ [12pt]
    \Delta^2_\mathcal{S}(k)&=\displaystyle\frac{k^3}{2\pi^2}P_\mathcal{S}(k)&=A_{\rm iso}(k_p)\left(\frac{k}{k_p}\right)^{n_{\rm iso}-1}\label{Eq:PS}
\end{eqnarray}
at a pivot scale $k_p=0.05~ \rm{Mpc}^{-1}$ and where $A_i ~(i={\rm ad, iso})$ and $n_i~({i=\rm ad, iso})$ are the amplitude and spectral index of the spectrum, respectively. If only CDM carries the isocurvature component, the amplitude of total matter isocurvature, $A_{\rm iso}$, is related to the CDI amplitude, $A_{\rm cdi}$, as 
$$A_{\rm cdi}(k_p)=f_c^2A_{\rm iso}(k_p),
$$where $f_c$ is the fraction of CDM in the total matter. Henceforth, we reserve \textit{iso} and \textit{cdi} to distinguish between the total matter and CDM isocurvature fluctuations, respectively. Our current best measurements of the CMB anisotropies 
from Plank satellite in Ref.~\cite{Planck:2018jri} place the values of $A_{\rm ad}$ and $n_{\rm ad}$ at \[A_{\rm ad}\approx 2.1 \times 10^{-9},\qquad n_{\rm ad} \approx 0.97.\]
Hence, adiabatic fluctuations are almost scale-free with a slight red-tilt signifying the slow-roll of the inflaton. In comparison, the uncorrelated scale-invariant CDM isocurvature modes have not been detected in the current data and are strongly constrained such that $A_{\rm cdi} \lesssim 0.04\, A_{\rm ad}(k_p)$ \cite{Planck:2018jri}. Hence, the scale-invariant isocurvature power 
should be roughly $\mathcal{O}(1\%)$ or less than adiabatic on large scales. Since current constraints are limited to large scales $\gtrsim \mathcal{O}(100)~$Mpc, it is possible to evade them by conjecturing isocurvature perturbations with a strongly blue spectral tilt ($n_{\rm iso}>1$) which can be completely negligible on large scales and be the dominant primordial inhomogeneity component on small length scales. 

The above expression in Eq.~(\ref{Eq:PS}) for the isocurvature spectrum assumes a pure power-law form and therefore lacks the spectral break that is generically expected and necessary for the observational viability of highly blue-tilted ($n_{\rm iso}\gtrsim 2.4$) spectrum for a spectator scalar field~\cite{Chung:2015tha}. In many axion-like models, the Peccei-Quinn radial field lies out of equilibrium (dynamically rolling) during inflation, producing strongly blue-tilted ($n_{\rm cdi}\sim 1-4$) isocurvature fluctuations ~\cite{Kasuya:2009up}. As the radial field relaxes to its vacuum during inflation, the axion spectrum transitions from this blue‐tilted regime to an approximately scale–invariant plateau, introducing a clear break in the spectrum. 

Other post‐inflationary mechanisms producing isocurvature fluctuations likewise imprint causal cutoffs in the isocurvature power: spatial variations in primordial black hole abundance~\cite{Afshordi:2003zb},
field fluctuations from symmetry breaking after inflation \cite{Enander:2017ogx}, and entropy modes generated during phase transitions \cite{Elor:2023xbz,Buckley:2024nen}, and all yield spectra rise steeply at low $k$ (with $n_{\rm iso}\sim4$) before rolling over at a characteristic scale set by causal physics. For simplicity, however, we will omit such a break in the isocurvature spectrum throughout this work and retain a single power‐law parameterization with a free spectral index and amplitude.

In the linear perturbation theory, the adiabatic and isocurvature components evolve autonomously. Consequently, the overall evolution of the perturbation results from a combined evolution of these two components independently. Thus, the total linear matter power spectrum in the presence of mixed adiabatic and uncorrelated CDM isocurvature  initial perturbations for a background $\Lambda$CDM cosmology may be written as\footnote{
In this paper, we do not consider correlated isocurvature fluctuations.
}
\begin{equation}
    P^{\rm mx}_{\rm \Lambda CDM}(k) = P^{\rm ad}_{\rm \Lambda CDM}(k) + P^{\rm cdi}_{\rm \Lambda CDM}(k).
\end{equation}

In cosmological analyses, it is often convenient to characterize the evolution of linear matter fluctuations on the scales of interest through a transfer function $T(k,z)$. For a given model $X$ and a choice of initial conditions (IC), we define
\begin{equation}
    T^{\rm IC}_{\rm X}(k) = \sqrt{\frac{P_{\rm X}^{\rm IC}(k)}{P_{\rm REF}^{\rm ad}(k)}}\label{eq:transfer_fn_defn}
\end{equation}
where $P_{\rm X}^{\rm IC}$ is the linear matter power spectrum of model $X$, and $P_{\rm REF}^{\rm ad}$ is the linear power spectrum of our reference (REF) $\Lambda$CDM cosmology model under standard adiabatic (ad) initial conditions\footnote{
In Eq.~\eqref{eq:transfer_fn_defn} one typically suppresses the explicit redshift dependence of the transfer function, $T_X(k)\equiv T_X(k,z)$, under the following assumptions: (i) the analysis is restricted to the linear regime, $\Delta^2(k,z)\ll1$, so that $P(k,z)=D^2(z)\,P(k,0)$ with a scale‐independent growth factor $D(z)$; (ii) all redshifts satisfy $z\ll z_{\rm eq}$, well into matter domination where residual radiation effects are negligible; (iii) the warm‐dark‐matter particles are non‐relativistic over the considered $z$–range and the modes lie outside the free‐streaming length scale; (iv) the background expansion history $H(z)$ is identical between the target and reference models. With these restrictions, it is easy to check via numerical convergence that $\bigl|T(k,z_1)/T(k,z_2)-1\bigr|\ll10^{-3}$ for all redshifts $z_1,z_2$ in the study.
}. We restrict our analysis to models containing a cosmological constant; hence, for brevity, we omit the ``$\Lambda$'', and refer to models such as $\Lambda$CDM simply as CDM, when the context is clear.

Thus, we write the transfer function of the matter power spectrum for mixed  adiabatic and uncorrelated CDI in a $\Lambda$CDM model as
\begin{eqnarray}
T^{\rm mx}_{\rm CDM}(k) 
&=& \sqrt{\displaystyle\frac{P^{\rm mx}_{\rm CDM}(k)}{P^{\rm ad}_{\rm CDM}(k)}}  \nonumber \\ [8pt]
&=& \sqrt{1 + \left(T^{\rm cdi}_{\rm CDM}(k)\right)^2} \nonumber \\ [8pt]
&=&  \sqrt{1 + \left(\displaystyle\frac{f_{\rm c}}{3}\sqrt{\alpha_{\rm cdi}}~R^{\rm cdi}_{\rm CDM}(k)\right)^2}\label{eq:Tmx_cdm}
\end{eqnarray}
where in the last line we have factored out the CDM contribution to the total matter isocurvature by defining \[ f_{\rm c} = \frac{\Omega_{\rm c}}{\Omega_{\rm m}}\] as the fraction of CDM energy density to the total matter, and \[\alpha_{\rm cdi}=\left. \frac{A_{\rm cdi}}{A_{\rm ad}}\right|_{k_p}\] as the ratio of the primordial amplitudes of the CDI and adiabatic fluctuations. Consequently, the quantity $R^{\rm cdi}_{\rm CDM}$ only encodes the spectral dependence of the blue-tilted CDI and the overall transfer function shape. From Ref.~\cite{Elena_Pierpaoli_1999}, the empirical expression for $R^{\rm cdi}_{\rm CDM}$ in Eq.~(\ref{eq:Tmx_cdm}), applicable over a wide range of scales from $k \in \sim(10^{-4},10^2) \;{\rm Mpc^{-1}}$ can be given as
\begin{equation} R^{\rm cdi}_{\rm CDM} (k) = \frac{T_{\rm cdi}(q)}{T_{\rm ad}(q)} \left(\frac{k}{0.05\,\rm{Mpc^{-1}}}\right)^{(n_{\rm cdi} -n_{\rm ad})/2}\label{eq:Tcdi}
\end{equation}where
\[ q = \frac{k}{\Omega_{\rm DM}h^2} {\rm Mpc}\] and $T_{\rm cdi(ad)}(q)$ are obtained after smoothing over the BAO oscillations\footnote{
For CDM isocurvature, the fitting functions are obtained as (\cite{Elena_Pierpaoli_1999})
\begin{align*}
T_{\text{cdi}} &= \left(1 + \frac{(40q)^2}{1 + 215q + (16q)^2 (1 + 0.5q)^{-1}} + (5.3q)^{8/5} \right)^{-5/4} \\
T_{\text{ad}} &= \frac{\ln(1 + 2.34q)}{2.34q} \left(1 + 3.89q + (16.1q)^2 + (5.46q)^3 + (6.71q)^4 \right)^{-0.25}
\end{align*}
}. 

Recent Planck-based analyses in Ref.~\cite{Planck:2018jri} constrain $\alpha_{\rm cdi}\lesssim 0.6$ at 2$\sigma$ for uncorrelated $n_{\rm cdi}$-free CDI models after marginalizing over the spectral index. The additional factor of $1/3$ in Eq.~(\ref{eq:Tmx_cdm}) is obtained by mapping superhorizon isocurvature modes to
curvature perturbations during the Matter-Dominated (MD) era due to the change in the effective sound speed. See footnote 1 in Ref.~\cite{Chung:2023syw}, and Ref.~\cite{Kodama:1986ud} for further details. Note that for redshifts $z\ll z_{\rm eq}$, the above transfer function is redshift-independent to a high degree of approximation. 

\begin{figure}[t]
    \centering
    \includegraphics[width=0.7\linewidth]{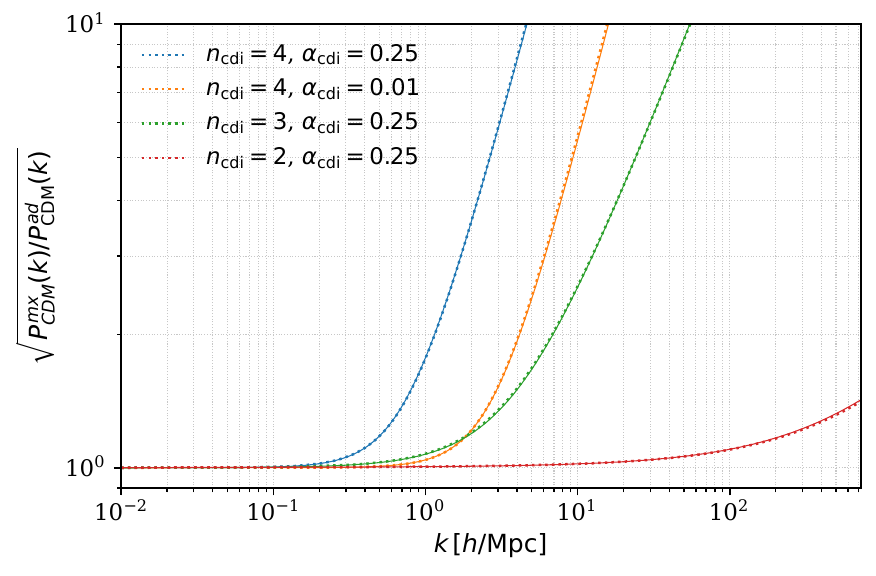}
    \caption{Matter transfer functions for the cosmological models with mixed adiabatic and CDI ICs, shown for various fiducial values of CDI spectral index $n_{\rm cdi}$ and amplitude $\alpha_{\rm cdi}$. The dotted curves are computed with a linear Boltzmann solver at $z=0$, and the solid curves denote the analytical approximation as taken from Eqs.~(\ref{eq:Tmx_cdm}) and (\ref{eq:Tcdi}).}
    \label{fig:T_CDMiso}
\end{figure}

In Fig.~\ref{fig:T_CDMiso}, we plot instances of total matter transfer functions for the cosmological model with mixed adiabatic and CDI ICs, illustrating how a blue-tilted CDI component influences the overall spectral shape. As discussed earlier, varying the amplitude and spectral index of the CDI component can enhance power beyond a specific scale.   

\subsection{Mixed warm and cold DM model}\label{Sec:WCDM}
Unlike CDM, which is non-relativistic with negligible pressure, during the radiation-dominated (RD) era, relativistic and semi-relativistic WDM particles travel cosmological distances, erasing density fluctuations below a “free-streaming” scale ($\lambda_{\rm FS}$). Consequently, pure WDM cosmologies lack adequate gravitational seeds for structure formation on scales below $\lambda_{\rm FS}$. Once they become non-relativistic, their evolution is nearly indistinguishable from CDM. For instance, WDM particles with masses $m_{\rm WDM}\gtrsim100$~eV can be safely assumed to be highly non-relativistic by the time of matter-domination of the background universe.  

We denote the transfer functions for the pure WDM models with arbitrary ICs as $T^{\rm IC}_{\rm WDM}$. Pure thermal WDM scenarios with adiabatic ICs have been studied extensively and their transfer functions are well described by the following analytical expression (\cite{PhysRevD.71.063534,Bode:2000gq,Viel:2011bk,PhysRevD.71.063534,Abazajian:2005gj})
\begin{equation}
    T^{\rm ad}_{\rm WDM}(k,z=0) = \left[1 + (\alpha k)^{2\nu}\right]^{-5/\nu}
\end{equation}
where $\nu$ and $\alpha$ are derived from empirical fits to the numerical data, largely obtained from linear Boltzmann solvers. The parameter $\alpha$ controls the cutoff/suppression scale of the transfer function such that for $T(k\sim 0.15 \alpha)\sim0.9$. Following recent fits derived in Ref.~\cite{Vogel:2022odl}, we write
\begin{equation}
\alpha (m_{\rm WDM}) = a \,
\left(\frac{m_{\rm WDM}}{1\ \mathrm{keV}}\right)^{b}
\left(\frac{\omega_{\rm WDM}}{0.12}\right)^{\eta}
\left(\frac{h}{0.6736}\right)^{\theta} 
\, h^{-1} \mathrm{Mpc},
\end{equation}where $a = 0.0437$, $b = - 1.188$, $\theta = 2.012$, $\eta = 0.2463$ and $\nu =1.049$ for spin-1/2 WDM particles, with a $10\%-$ accuracy on scales $k\lesssim 2\alpha ~\rm{h/Mpc}$ for WDM masses $\gtrsim \mathcal{O}(0.5)$ keV.

In the mixed warm and cold DM cosmological model, henceforth referred to as WCDM, the fraction of warm DM component is given as
\begin{equation}
    f_{\rm WDM} = \frac{\Omega_{\rm WDM}}{\Omega_{{\rm DM}}}\label{eq:fwdm}
\end{equation}where $\Omega_{{\rm DM}}$ is the energy density fraction from the combined contribution of cold and warm DM components. In mixed WCDM models, one usually gives a lower limit constraint on the WDM masses $m_{\rm WDM}$ for varying WDM fractions $0<f_{\rm WDM}\leq1$.

In the WCDM model, modes with wavelength below the WDM free-streaming length, $\lambda_{\rm FS}$, are seeded almost entirely by the CDM component, since WDM's residual thermal velocities suppress its clustering on these scales. Consequently, for $k\gg k_{\rm FS}\sim \lambda_{\rm FS}^{-1}$, the matter power spectrum asymptotes to a plateau set by pure CDM, and the total transfer function $T^{\rm ad}_{\rm WCDM}$ exhibits a constant value at high $k$ \cite{Boyarsky:2008xj}. A convenient approximation for the transfer function of the WCDM model is often given as
\begin{equation}
    T^{\rm ad}_{\rm WCDM}(k) \approx f_{\rm WDM}T^{\rm ad}_{\rm WDM}(k) + \left(1-f_{\rm WDM}\right)  \label{eq:Tk_WCDM_0}   
\end{equation}which holds for $k\lesssim k_w$ such that $T^{\rm ad}_{\rm WCDM}(k)\gtrsim 1-f_{\rm WDM}$.

For $k\gtrsim k_w$, the approximation in Eq.~(\ref{eq:Tk_WCDM_0}) breaks down as the first term becomes subdominant. As shown in Ref.~\cite{Boyarsky:2008xj}, for $f_{\rm WDM}\gtrsim 0.1$, the true high-$k$ plateau lies well below the naive estimate \( 1 - f_{\text{WDM}} \). Physically, this strong suppression arises because (i)~WDM free streaming shallowens the gravitational wells initially seeded by CDM, and (ii)~CDM perturbations then grow more slowly in these weakened potentials. Furthermore, the suppression is not simply a direct subtraction effect, since the full evolution of $\delta_{\rm CDM}$ and $\delta_{\rm WDM}$ obey coupled linear differential equations. As WDM begins to accrete into the CDM-generated potential wells, its own transfer function is ``dragged up'' relative to a pure WDM model, but its inefficient clustering still modifies the overall initial conditions for structure growth. The net result is a smooth interpolation of the $\delta_m$  transfer function from the WDM cutoff scale to a diminished CDM plateau, rather than a sharp two-term step.

\begin{figure}[ht]
    \centering
    \includegraphics[width=0.5\linewidth]{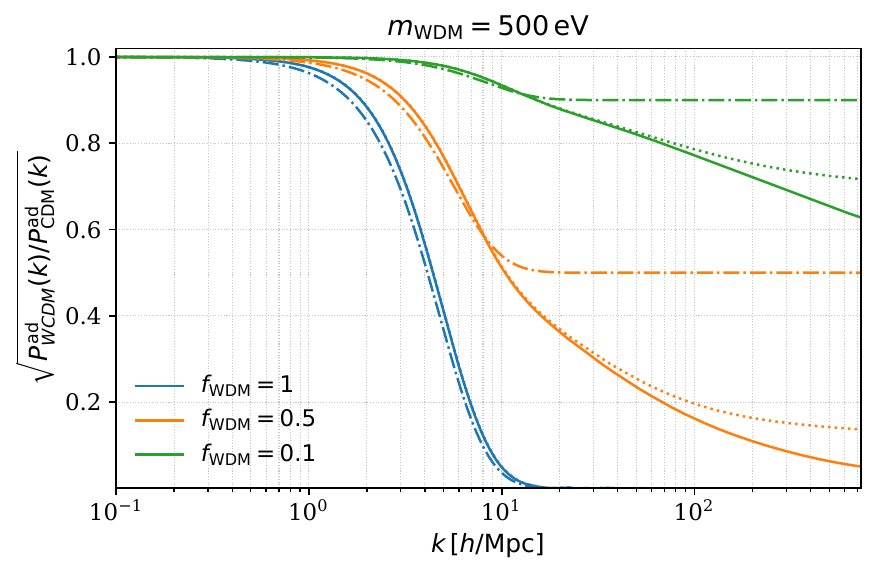}~~    \includegraphics[width=0.5\linewidth]{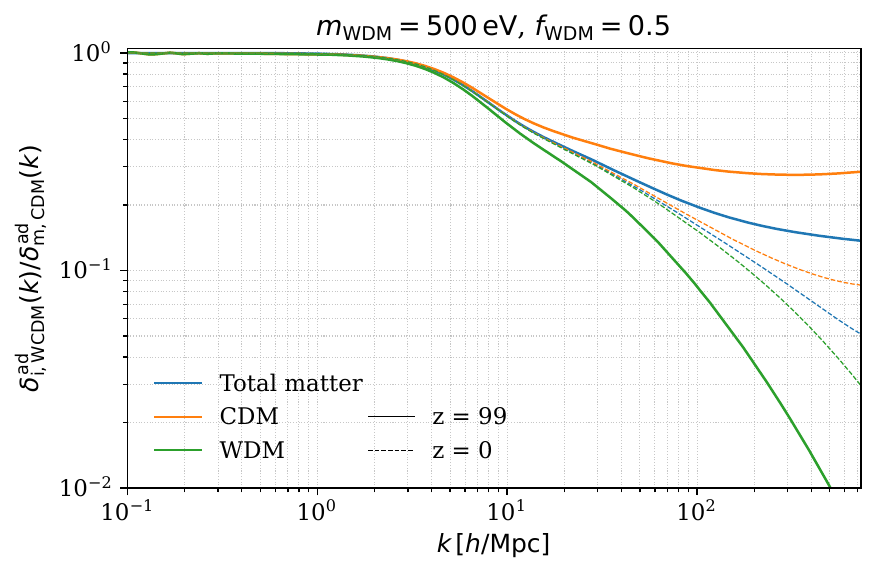}
    \caption{\textbf{Left}: Total matter transfer functions for the WCDM model with adiabatic IC for $m_{\rm WDM}=500$ eV, shown for three values of the WDM fraction $f_{\rm WDM}$. Solid and dotted curves are computed with a linear Boltzmann solver at $z=0$ and $z=99$, respectively. Dash-dotted curves denote the analytical approximation of Eq.~(\ref{eq:Tk_WCDM_0}). \textbf{Right}: Transfer‐function ratios of the WCDM model ($f_{\rm WDM}=0.5$) with adiabatic IC for total matter, CDM, and WDM, each normalized to the total-matter transfer in the reference $\Lambda$CDM cosmology, shown at $z=0$ (solid) and $z=99$ (dashed).}
    \label{fig:T_WCDM}
\end{figure}

Fig.~\ref{fig:T_WCDM} shows the individual $\delta_{\rm WDM(CDM)}$ and total $\delta_m$ transfer functions computed with the linear Boltzmann solver, \textsf{CLASS} (\cite{Blas:2011rf}), in a WCDM model for a few $f_{\rm WDM}$ fractions at a fixed WDM mass of $500$ eV with only adiabatic IC. One clearly sees (i) the WDM and CDM curves converging over $k$-range ($\alpha k\lesssim 0.5 $), and (ii) the final mixed-model plateau lying below $1-f_{\rm WDM}$, in agreement with the above discussion. Unlike in a pure WDM model, the WCDM transfer function retains a non-negligible dependence on redshift for scales $k\gtrsim k_{\rm FS}$. This arises because, for modes with $k\gtrsim k_{\rm FS}$, the WDM perturbations remain suppressed relative to CDM ($\delta_{\rm WDM}(z) < \delta_{\rm CDM}(z)$), so that the total matter clustering and hence $T^{\rm ad}_{\rm WCDM}$ evolves as the two components grow at different rates. At lower redshifts, the high-$k$ CDM plateau is visibly diminished compared to its value at early times. As demonstrated in Ref.~\cite{Boyarsky:2008xj}, this $z$-dependence appears only for scales where the WDM contrast lags behind CDM, which is applicable for $k>k_{\rm FS}$. As observed in the right panel of the Fig.~\ref{fig:T_WCDM}, while the CDM drags the WDM components along during the redshift evolution, the free-streaming of the WDM suppresses the CDM transfer function at lower redshifts compared to its value at early times.

\begin{figure}
    \centering
    \includegraphics[width=0.7\linewidth]{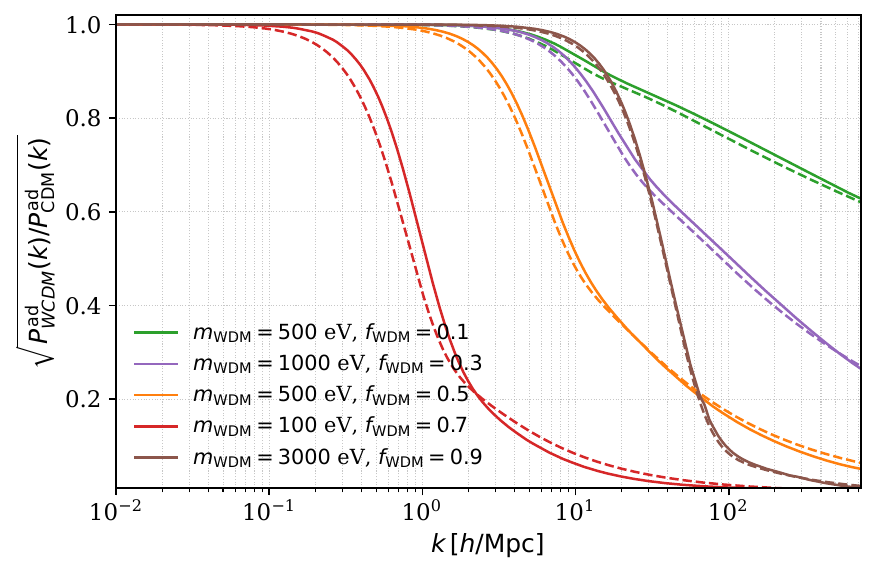}
    \caption{Linear matter transfer functions at $z=0$ for the WCDM models with adiabatic ICs and varying choice of WDM mass $m_{\rm WDM}$ and fraction $f_{\rm WDM}$. The solid curves are obtained from a linear Boltzmann solver, while the dashed curves represent our approximation based on the expression in Eq.~(\ref{eq:Tk_mod_WCDM}).}
    \label{fig:modifiedTWCDM}
\end{figure}

The trends visible in Fig.~\ref{fig:T_WCDM} demonstrate that the simple interpolation of Eq.~\eqref{eq:Tk_WCDM_0} is inadequate.  Existing fitting functions in the literature adopt three or four free parameters $(\alpha,~\beta,~\gamma,~\delta)$ that are agnostic to the underlying warm-component mass and fraction \cite{Murgia:2017lwo,Hooper:2022byl}.  Although such fits are convenient for simulation and MCMC scans, they obscure the physical dependence on $m_{\rm WDM}$ and $f_{\rm WDM}$.

Guided by the trends in Fig.~\ref{fig:T_WCDM}, we propose the following empirical refinement of Eq.~(\ref{eq:Tk_WCDM_0}) by introducing a correction factor to the CDM:
\begin{equation}
    T^{\rm ad}_{\rm WCDM}(k,z=0)\;\simeq\;
    f_{\rm WDM}\,T^{\rm ad}_{\rm WDM}(k)
    \;+\;
    \bigl(1-f_{\rm WDM}\bigr)C(k),
    \label{eq:Tk_mod_WCDM}
\end{equation}where
\begin{equation}
    C(k) = \left[T^{\rm ad}_{\rm WDM}(k)\right]^{b f_{\rm WDM}}
\end{equation} and we find $b\approx 0.1$ as a suitable choice.

Eq.~\eqref{eq:Tk_mod_WCDM} reproduces the intermediate-scale leveling where the WDM drags on to CDM, thereby softening the CDM potential wells, and preserves an explicit and physically interpretable dependence on $m_{\rm WDM}$ and $f_{\rm WDM}$.

Tests with a linear Boltzmann solver in Fig.~\ref{fig:modifiedTWCDM} show that Eq.~\eqref{eq:Tk_mod_WCDM} tracks the full numerical transfer function to within a few percent for a wide range of WDM mass and fraction, with the deviations at intermediate-scales partly limited by the accuracy of the $T^{\rm ad}_{\rm WDM}(k)$ fitting form given in Eq.~(\ref{eq:Tk_WCDM_0}).

\section{WCDM with CDI: linear analysis}\label{sec:WCDM+CDI}
In the preceding section, we discussed that thermal-WDM induces a pronounced suppression of small-scale (below $\lambda_{\rm FS}$) power due to free–streaming (Sec.~\ref{Sec:WCDM}), while uncorrelated CDM isocurvature can, if strongly blue‐tilted, inject additional power precisely on those same scales (Sec.~\ref{sec:CDMiso}). In this section, we study a cosmology in which both effects coexist: a mixed WCDM+CDI model. By tuning $n_{\rm cdi}$ and $\alpha_{\rm cdi}$, one can therefore restore (or even enhance) the small–scale amplitude relative to the pure WDM case, effectively compensating WDM suppression with CDI power. All results presented in this section are computed within linear perturbation theory using a Boltzmann solver \textsf{CLASS}.

We shall assume that only the CDM component carries isocurvature fluctuations. The conditions for survival of the isocurvature fluctuations require that its carrier doesn’t thermalize with the primordial thermal plasma or carries a non-zero conserved charge \cite{Weinberg:2004kf} and is stable (does not decay) to affect cosmological fluid initial conditions at late time. These conditions are readily fulfilled by non-thermal DM candidates such as axions. See Refs.~\cite{Kawasaki:2013ae,Marsh:2015xka} for a review on axion cosmology. We will restrict our analysis to uncorrelated CDI. We are mostly interested in tiny CDM fractions, $1-f_{\rm WDM}\lesssim 0.1$, such that WDM makes up majority of the DM energy density. We will study and emphasize that CDM with even $1\%$ fraction with large blue-tilted uncorrelated isocurvature can compensate the effect of WDM particles.

\subsection{Joint CMB and BAO constraint}\label{sec:cmb_bao}
CMB anisotropies provide some of the most stringent limits on uncorrelated CDM isocurvature because isocurvature modes leave imprints across the temperature and polarization spectra.  Unlike adiabatic fluctuations, isocurvature perturbations shift the acoustic oscillation phase and modify the relative heights of the TT peaks, producing features in TE and EE that cannot be mimicked by changes in standard cosmological parameters.  On the largest scales, they alter the early Integrated Sachs–Wolfe contribution to the low-$\ell$ TT plateau, while in polarization the TE correlation even changes sign for an entropy mode.  The combination of these effects, measured from CMB with sub-percent precision, therefore anchors the CDI fraction very strongly at the large scales.

As discussed in Sec.~\ref{sec:CDMiso}, Planck’s measurements constrain uncorrelated $n_{\rm cdi}$-free CDI to have $\alpha_{\rm cdi}\lesssim0.6$ at the pivot scale $k_p=0.05$~Mpc$^{-1}$~\cite{Planck:2018jri}. Complementing this, the recent ACT DR6 analysis, probing down to smaller scales, finds $A_{\rm cdi}\lesssim2~A_{\rm ad}$ at $k=0.1$~Mpc$^{-1}$, with the CDI spectral index within the range $1.0\lesssim n_{\rm cdi}\lesssim3.0$ at $95\%$ C.L. \cite{ACT:2025tim}. Both of these studies also incorporate late-time geometric information from BAO measurements by BOSS \cite{BOSS:2016wmc} and 6dFGS \cite{Beutler:2011hx} in the Planck analysis and by DESI \cite{DESI:2024uvr,DESI:2024lzq,DESI:2024mwx} in the ACT analysis to tighten their bounds on the isocurvature contribution.

Below, we present results from Markov‑chain Monte‑Carlo (MCMC) sampling with \textsf{Cobaya} \cite{Torrado:2020dgo}, jointly fitting CMB and BAO likelihoods to the mixed WCDM+CDI model. Our objective is to map the joint WDM and isocurvature parameter region that remains consistent with large‑scale data. For each MCMC evaluation, we run four parallel chains starting from a precomputed covariance matrix. The run is deemed satisfactorily converged when the Gelman-Rubic criterion $R-1<0.01$. We utilize \textsf{getdist} to derive constraints from the MCMC chains. We consider the CMB anisotropic likelihood from Planck. This includes planck2018 lowl.TT and lowl.EE \textsf{SRoll2}   \cite{Pagano:2019tci} likelihoods to anchor the large-angle TT and EE spectra ($\ell\lesssim 30$), matching the Planck low-multipole treatment. We add Planck's baseline nuisance-marginalized high-multipole TT/TE/EE measurements (\textsf{Plik}) up to $\ell_{\max} =2500$ with a floating calibration parameter $A_{\rm Planck}$ \cite{Planck:2018lkk}. For the CMB lensing potential likelihood, we consider Planck PR3 bandpowers within the range $8\leq L\leq 400$ \cite{Planck:2018lbu}. Finally, we add the BAO measurements from DESI Y1 observations \cite{DESI:2024uvr,DESI:2024lzq,DESI:2024mwx}.

We work within a flat $\Lambda$WCDM framework with massless neutrinos. We consider mixed DM models for two fiducial warm fractions fixed to $f_{\mathrm{WDM}} = 0.99$ and $0.5$. In the former, the scenario is effectively a ``nearly pure'' WDM cosmology, with only a trace CDM component. This fiducial choice lets us probe how even a trace CDM admixture through its isocurvature can modify an otherwise WDM‑dominated cosmology. The latter allows us to identify the constraining power of large-scale data from CMB and BAO as the warmness criterion is relaxed considerably.
We assume that all species (matter and radiation) carry usual nearly scale-invariant adiabatic fluctuations, while only the CDM carries the additional isocurvature fluctuations. We generate the linear power spectrum of the CMB anisotropies and matter with \textsf{CLASS}. Because existing nonlinear prescriptions (e.g. \textsf{HMcode} \cite{Mead:2020vgs}, \textsf{Halofit} \cite{Takahashi:2012em}) are not calibrated for strongly blue CDI and are found to underpredict power on even linear scales for $n_{\rm{cdi}}\gtrsim 3$ (see Ref.~\cite{Chung:2023syw}), we restrict to using the linear matter power spectrum for lensing calculations, rendering our constraints deliberately conservative.

In \textsf{CLASS}, the primordial CDI is specified by two parameters: the spectral index $n_{\rm cdi}$ and the amplitude $f_{\rm cdi}$, where \[f_{\rm cdi} \equiv \sqrt{\alpha_{\rm cdi}}= \sqrt{A_{\rm cdi}/A_{\rm ad}}\] at the pivot scale $k_p$. However, because CDM isocurvature sources only the CDM component, its impact on the total matter perturbation is suppressed by the CDM fraction $f_c=\omega_c/\omega_m$. Consequently, the physically invariant amplitude of the matter isocurvature mode is given by the product
\begin{equation}
    f_{\rm iso} = \sqrt{\alpha_{\rm iso}}= f_c f_{\rm cdi}.\label{eq:fiso}
\end{equation}Holding $f_{\rm iso}$ fixed, while keeping all other background densities and the CDI spectral shape unchanged, guarantees that the CDI-only contribution to both the matter power spectrum and the CMB multipoles remains the  same. 

Because the CMB shows no statistically significant CDI signal (the posterior for its amplitude piles up at the physical lower bound $A_{\rm cdi}=0$), parameter estimates are sensitive to how the CDI sector is parametrized. To avoid prior-volume and boundary effects, we forgo the pair ${n_{\rm cdi},\,f_{\rm cdi}}$ and instead sample the CDI power spectrum amplitude directly at two wavenumbers,
$$
k_1=0.002\ \text{Mpc}^{-1}, \qquad
k_2=0.05\ \text{Mpc}^{-1}.
$$ as adopted by several previous CMB analyses. For example, see Ref.~\cite{Planck:2018jri}. Our choice of $k_2$ is lower than the $0.1\, \rm{Mpc}^{-1}$ pivot adopted by the Planck and ACT analyses \cite{Planck:2018jri,ACT:2025tim}. This lets us quote the CDI amplitude at the conventional pivot $k_p$ without introducing extra Jacobian factors when transforming variables.

Thus, our full set of sampled parameters is
$$
\Bigl\{\,
\omega_b,\,
\omega_c,\,
\theta_s,\,
\mathcal{P}_{\mathcal{RR}}^{(1)},\,
\mathcal{P}_{\mathcal{RR}}^{(2)},\,
\tau_{\mathrm{reio}},\,
\log_{10}(m_{\mathrm{WDM}}),\,
\mathcal{P}_{\mathcal{II}}^{(1)},\,
\alpha_{\mathrm{iso}}
\Bigr\},
$$
where $\mathcal{P}_{\mathcal{RR}}^{(i)}$ and $\mathcal{P}_{\mathcal{II}}^{(i)}$ are the adiabatic and CDI power spectrum amplitudes evaluated at $k_i$ ($i=1,2$). The CDI amplitude at the second scale is derived, not independent:
$$
\mathcal{P}_{\mathcal{II}}^{(2)}
= \mathcal{P}_{\mathcal{RR}}^{(2)}
\alpha_{\mathrm{iso}}f^{-2}_c.
$$
with $f_c$ being the CDM fraction. We derive the spectral tilt $n_{\rm cdi}$ as follows:
$$
n_{\rm cdi} = 1 + \frac{\log (\mathcal{P}_{\mathcal{II}}^{(2)}/\mathcal{P}_{\mathcal{II}}^{(1)})}{\log (k_2/k_1)}.
$$

\begin{figure}[ht]
    \centering
    \includegraphics[width=0.75\linewidth]{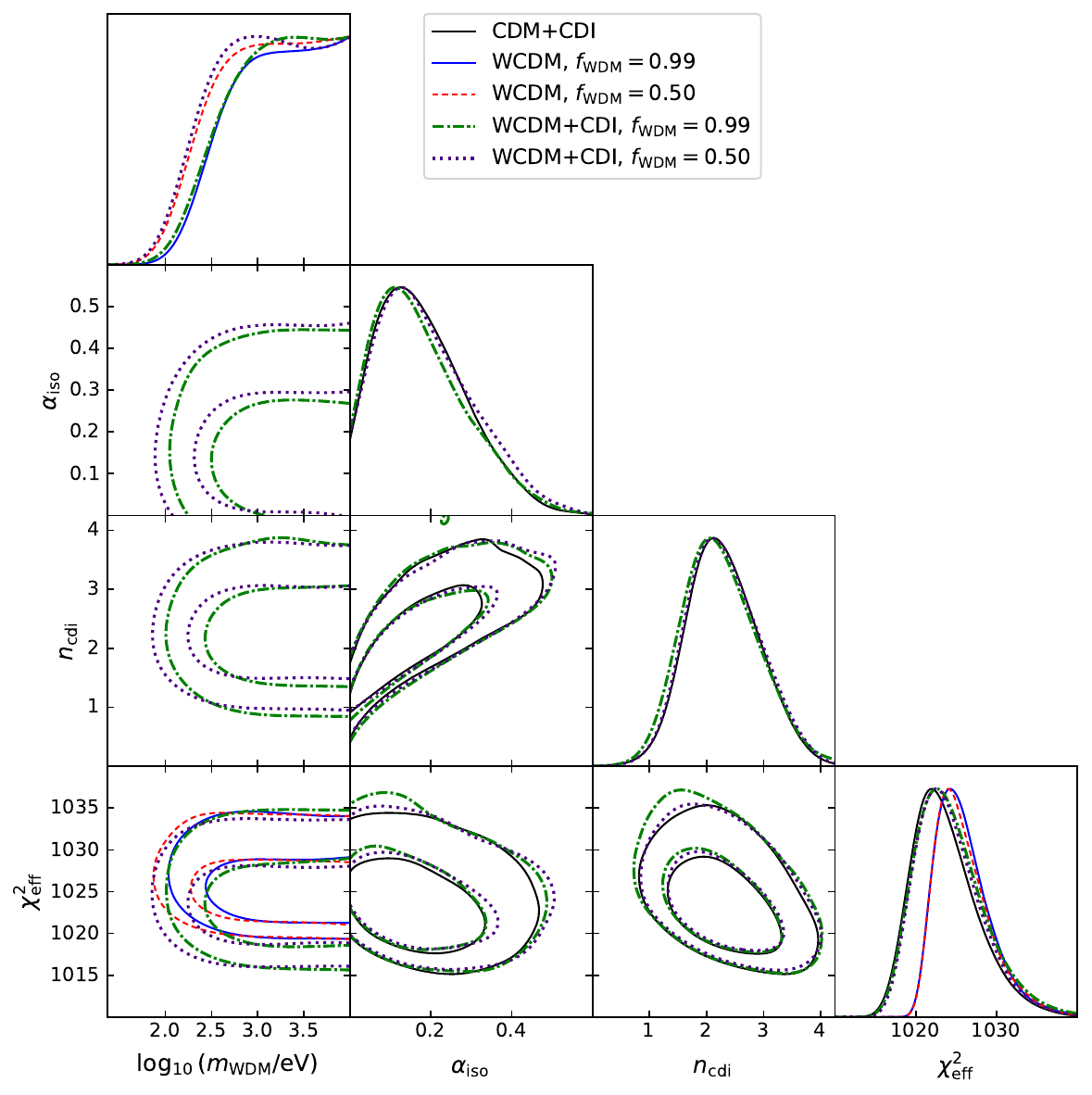}
    \caption{Marginalized posterior showing constraints on parameters $\log_{10}(m_{\rm WDM}),~n_{\rm cdi}$ and $\alpha_{\rm iso}$ for the WCDM models, both with and without CDI, obtained using a joint analysis of CMB (Planck) and BAO (DESI) datasets. The effective $\chi^2$ from the MCMC sampling is also shown. We consider two sets of warm-fractions, $0.5$ and $0.99$ for with and without CDI models. For comparison, we also show a CDM+CDI model with no WDM.}
    \label{fig:planck_act}
\end{figure}

In Fig.~\ref{fig:planck_act}, we present the constraints on parameters $\log_{10}(m_{\rm WDM}),~n_{\rm cdi}$ and $\alpha_{\rm iso}$ for the WCDM models with and without CDI. For comparison, we also present CDI constraints from the vanilla CDM+CDI model with no WDM. The overlaid posteriors for $\log_{10}(m_{\rm WDM})$ with and without CDI are essentially indistinguishable.  In other words, marginalizing over a modest CDI component does not shift or broaden the WDM mass constraints, coming from Planck CMB data and DESI BAO measurements, appreciably. This reflects the fact that these datasets probe scales ($\ell\lesssim2500$, $k\lesssim0.1\,h/\mathrm{Mpc}$) where CDI‐induced power enhancements are too small to alter the inference of $m_{\rm WDM}$.

Although allowing for CDI has almost no impact on the derived lower bound for $\log_{10}(m_{\rm WDM})$ from Planck+DESI, it yields a well‐defined 1$\sigma$ and 2$\sigma$ credible regions in the $(\alpha_{\rm iso},n_{\rm cdi})$ plane. These allowed ranges will be sampled in our subsequent small‐scale analyses, ensuring that any CDI‐driven enhancement/compensation of power at $k\gtrsim1\,h/\mathrm{Mpc}$ remains fully consistent with the large‐scale CMB and BAO constraints.

Our recovered CDI amplitude and spectral tilt are consistent with the bounds quoted by the Planck and ACT collaborations \cite{Planck:2018jri, ACT:2025tim}. Allowing a CDI contribution lowers the effective $\chi^{2}$ chiefly through a better fit to the CMB-lensing likelihood. However, unlike those earlier studies, our analysis (i) neglects the nonlinear correction to the lensing and (ii) fixes the summed neutrino mass to zero.

Figure \ref{fig:planck_act} shows that the marginalized posterior for the amplitude, $\alpha_{\rm iso}$, peaks near $0.1$ and is almost independent of the CDM fraction $f_{c}$. Consequently, in the mixed WCDM+CDI scenarios with very little CDM ($f_{c}\sim0.01$), the CDI amplitude at the pivot scale $k_{p}=0.05\ \mathrm{Mpc}^{-1}$ can reach $\alpha_{\rm cdi}\sim100$, even though the total matter isocurvature is extremely small. Thus, such large CDI amplitudes can remain hidden even in small-scale spectral distortion measurements from COBE/FIRAS \cite{Fixsen:1996nj} and may become marginally detectable in future experiments such as PIXIE \cite{Kogut:2011xw}. In the next section, we briefly discuss the implications of such a large CDI amplitude.

\subsection{Linear matter power spectrum}\label{sec:WCDM+CDI:linear_mPK}
In the linear theory of perturbations, the matter power spectrum in the WCDM model for the mixed adiabatic and uncorrelated CDI initial conditions can be written as a linear superposition of the adiabatic and isocurvature components:
\begin{equation}
    P_{\rm WCDM}^{\rm mx}(k) = P_{\rm WCDM}^{\rm ad}(k) + P_{\rm WCDM}^{\rm cdi}(k).
\end{equation}
Hence, we can approximate the total matter transfer function as
\begin{align}
    T_{\rm WCDM}^{\rm mx}(k) \approx \sqrt{ \left(f_{\rm WDM}T^{\rm ad}_{\rm WDM}(k) + \left(1-f_{\rm WDM}\right)C(k)\right)^2 + \left( \frac{f_c f_{\rm cdi}}{3} ~R^{\rm cdi}_{\rm CDM}(k)C(k)\right)^2}
 \label{eq:Tk_WCDM_CDI}
\end{align}
where we take $f_c=(1-f_{\rm WDM})\omega_{\rm DM}/\omega_{\rm m}$, $f_{\rm cdi}=\sqrt{\alpha_{\rm cdi}}$ and the CDM correction factor is given as
\begin{equation}
    C(k) = \left[T^{\rm ad}_{\rm WDM}(k)\right]^{b f_{\rm WDM}}.
\end{equation}Through empirical fitting, we find $b\approx 0.125$ as a suitable choice for the WCDM model with mixed ICs. 

\begin{figure}[t]
    \centering
    \includegraphics[width=0.5\linewidth]{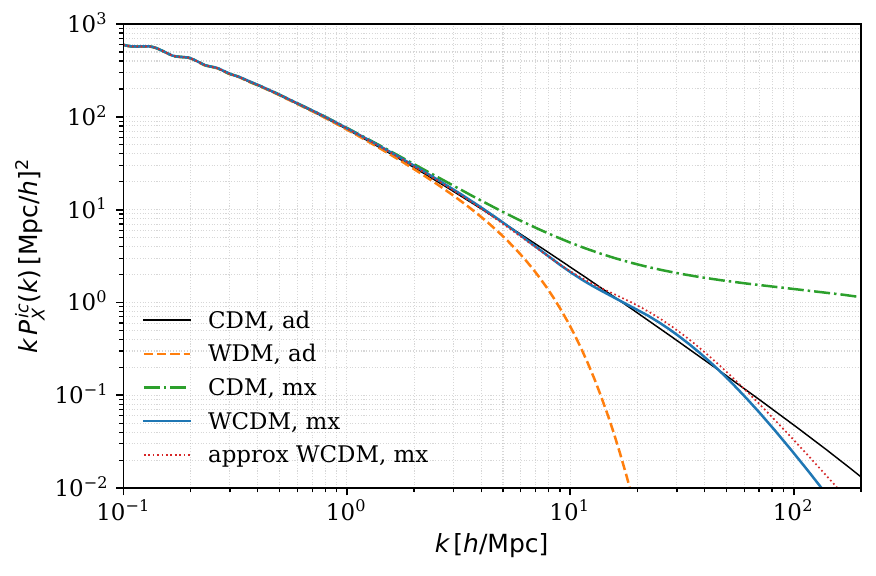}~
    \includegraphics[width=0.5\linewidth]{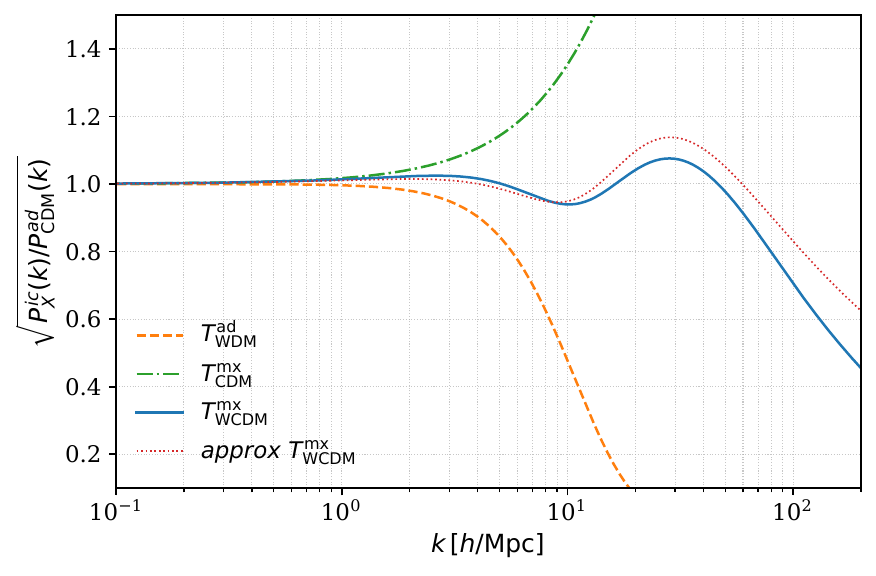}
\caption{We highlight the shape of the linear matter power spectrum (on left) and transfer function (on right) computed with CLASS at a redshift of $z=0$ for the WCDM model (solid blue curve) with mixed adiabatic and CDI ICs. The WDM component is set with $m_{\rm WDM}=1000~\rm{eV}$ and $f_{\rm WDM}=0.99$, while the CDI is initialized as $n_{\rm cdi}=2.83$ and $f_{\rm cdi}=0.293/(1-f_{\rm WDM})$. The dotted red curve is our semi-analytic approximation in Eq.~(\ref{eq:Tk_WCDM_CDI}) matching the numerical result to better than 10-percent over an extended range of scales. For comparison, we highlight the suppression in the power spectrum for a pure WDM model with adiabatic ICs for the same $m_{\rm WDM}=1000~\rm{eV}$ through the dashed orange curve, while the dot-dashed green curve corresponds to a standard CDM model with the same mixed ICs ($n_{\rm cdi}=2.83$ and $f_{\rm cdi}=0.293$) but no warm component. Together, they illustrate the compensation of the small-scale damping from WDM free-streaming by a suitable scale-dependent boost from the blue-tilted CDI model.}
    \label{fig:TWCDMiso}
\end{figure}

In Fig.~\ref{fig:TWCDMiso}, we illustrate the compensation of the small-scale suppression from WDM free-streaming by a suitable scale-dependent boost from the blue-tilted CDI fluctuations. The orange dashed curve represents the transfer function for a pure WDM model with adiabatic ICs for $m_{\rm WDM}=1000~\rm{eV}$. In our fiducial WCDM model with mixed ICs, we consider a $1\%$ CDM contribution to the total DM, and include CDI initialized with  $n_{\rm cdi}=2.83$ and $f_{\rm cdi}=0.293/(1-f_{\rm WDM})$ which lie within the $2\sigma$ allowed region in Fig.~\ref{fig:planck_act}. The resulting linear transfer function is plotted as a solid blue curve. The dotted red curve is our analytic approximation from Eq.~(\ref{eq:Tk_WCDM_CDI}) matching the numerical result up to $10$-percent accuracy over an extended range of scales ($k\lesssim k_{\rm FS}$). For comparison, we also plot the dot-dashed green curve, which corresponds to a standard CDM model with the same mixed ICs ($n_{\rm cdi}=2.83$ and $f_{\rm cdi}=0.293$) but no warm component ($f_{\rm WDM}=0$). The matter transfer function for our fiducial model, represented by the solid blue curve, exhibits a nearly order unity amplitude up to the free-streaming scale ($\sim \mathcal{O}(1)/\alpha$) of a $1$ keV WDM.

\begin{figure}[t]
    \centering
    \includegraphics[width=0.5\linewidth]{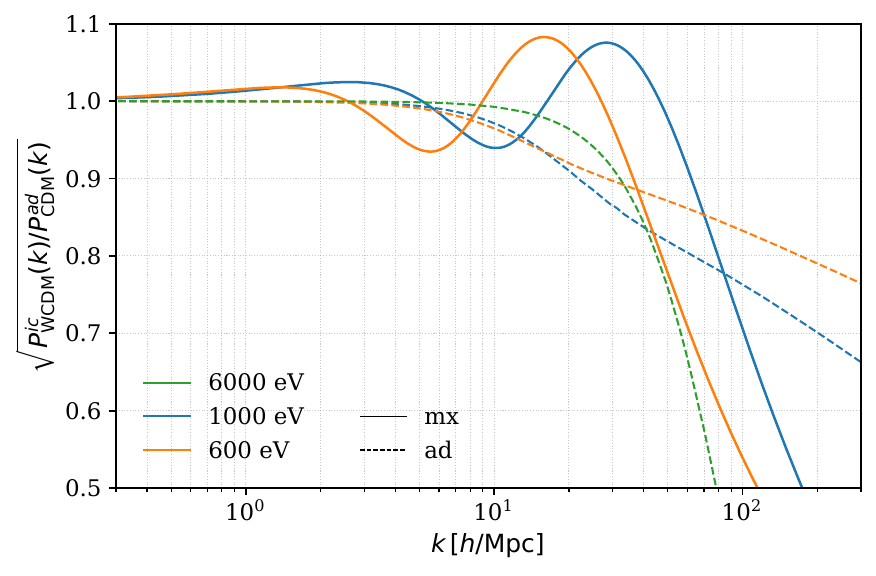}~\includegraphics[width=0.52\linewidth]{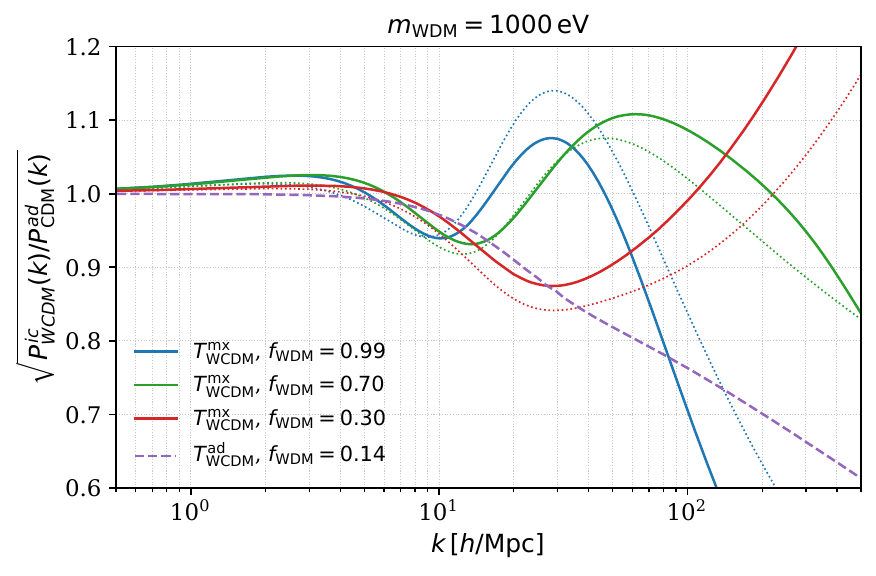}
    \caption{Linear matter transfer functions for various WCDM models. \textbf{Left:} The dashed blue (orange) curves correspond to WCDM models with 
$m_{\rm WDM}=1000~(600)$~eV and lower warm fractions 
$f_{\rm WDM}=0.14~(0.08)$ under purely adiabatic initial conditions, representing the approximate lower bounds on the WCDM parameter space. For comparison, the dashed green curve shows a pure WDM scenario with 
$m_{\rm WDM}=6$~keV and $f_{\rm WDM}=1$, illustrating the $2\sigma$ lower limit on the WDM mass for fully warm dark matter. The solid blue (orange) curves are obtained for 
$f_{\rm WDM}=0.99$ with $m_{\rm WDM}=1000~(600)$~eV and include mixed initial conditions specified by 
$n_{\rm cdi}=2.83~(3.0)$ and $f_{\rm cdi}(1-f_{\rm WDM})=0.293~(0.313)$. These demonstrate that introducing a correlated CDI component can produce transfer functions similar to adiabatic models with substantially higher warm fractions, effectively mimicking the suppression otherwise achieved only with smaller $f_{\rm WDM}$. \textbf{Right:} Further, we plot $T^{\rm mx}_{\rm WCDM}$ as solid curves for $m_{\rm WDM}=1$ keV with varying warm fractions $f_{\rm WDM}=0.99~(0.70)~(0.30)$ under mixed ICs specified by $n_{\rm cdi}=2.83~(2.68)~(2.4)$ and $f_{\rm cdi}(1-f_{\rm WDM})=0.293~(0.33)~(0.32)$. The dotted curves are obtained from our approximate formula in Eq.~(\ref{eq:Tk_mod_WCDM}). The dashed purple curve represents the current allowed limit on the warm fraction for a $1$ keV WDM with adiabatic-only IC.
}
    \label{fig:TWCDMiso-2and3}
\end{figure}

\begin{figure}
    \centering
    \includegraphics[width=0.7\linewidth]{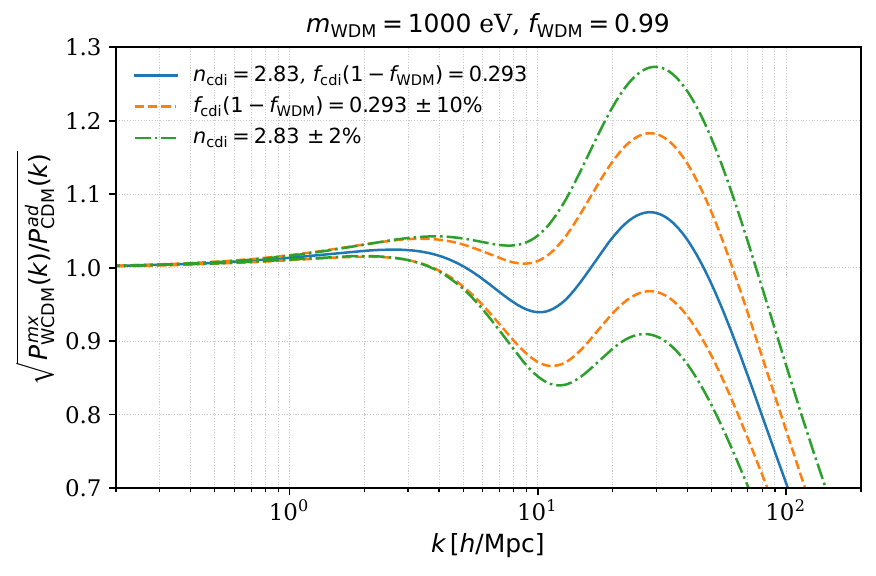}
\caption{Impact of CDI parameter variations on the mixed-IC WCDM transfer function for 
$m_{\rm WDM}=1000\,$eV and $f_{\rm WDM}=0.99$.  
The solid blue curve is the fiducial model 
($n_{\rm cdi}=2.83$, $f_{\rm cdi}(1-f_{\rm WDM})=0.293$).  
The dashed orange curves show a $\pm10\%$ shift in the CDI amplitude, producing an almost uniform 
offset in the high-$k$ features.  
The dash–dot green curves correspond to a 
$\pm2\%$ variation in the spectral index, which mainly alters the relative height of the 
oscillatory bumps. This demonstrates that the CDI amplitude controls the overall normalization of 
$T^{\rm mx}_{\rm WCDM}(k)$, whereas the spectral index governs the shape of the oscillatory feature and the high-k power-law scaling.
}
    \label{fig:TWCDMiso-4}
\end{figure}

To further illustrate how the addition of an uncorrelated CDI mode to the WCDM model can mimic the transfer function normally achieved only by lowering the warm fraction, in Fig.~\ref{fig:TWCDMiso-2and3} (left) we show the linear matter transfer functions for two sets of adiabatic-only WCDM models plotted as dashed blue (orange) curves with $m_{\rm WDM}=1000~(600)$ eV and $f_{\rm WDM}=0.14~(0.08)$ respectively, that bracket the approximate lower limits of the WCDM parameter space as obtained from recent high-resolution high-redshift Lyman-$\alpha$ based studies in Refs.~\cite{Irsic:2023equ,Garcia-Gallego:2025kiw}. For reference, the dashed green line denotes a pure $6$ keV WDM model (\(f_{\rm WDM}=1\)) which sits at the boundary of the \(2\sigma\) mass bound. By contrast, the solid blue and orange curves computed for the same masses but with $f_{\rm WDM}=0.99$ and mixed ICs lie almost on top of the low‐fraction adiabatic cases, demonstrating that a suitably uncorrelated CDI component ``may'' compensate the small‐scale damping in intermediate $k$-range even when the warm fraction is near unity. In the right panel of Fig.~\ref{fig:TWCDMiso-2and3}, we extend this comparison for $m_{\rm WDM}=1$ keV with three different values of $f_{\rm WDM}=0.99,~0.70,~0.30$ (solid curves, mixed ICs as indicated) confirming the viability of the compensating mechanism across a wide range of warm fractions. We also overlay the corresponding predictions of our semi‐analytic fit from Eq.~\eqref{eq:Tk_mod_WCDM} as dotted curves.

Although introducing a blue-tilted CDM-isocurvature component can partially compensate the free-streaming suppression of WDM power over an intermediate $k$-range, the inherently distinct scale-dependences of the two transfer functions prevent a perfect cancellation. As a result, we emphasize that our combined mixed ICs WCDM transfer function retains residual oscillatory features as seen in Figs.~\ref{fig:TWCDMiso} and \ref{fig:TWCDMiso-2and3}. To assess how this partial compensation manifests at nonlinear scales, we turn to the $N$-body simulations in Sec.~\ref{sec:WCDM_NL} and analyze the corresponding nonlinear transfer functions.

In Fig.~\ref{fig:TWCDMiso-4}, we explore the shape and overall normalization of these residuals in the linear transfer functions by varying the CDI spectral index and amplitude around our fiducial case. Adjusting the CDI amplitude merely rescales the overall normalization of $T^{\rm mx}_{\rm WCDM}(k)$, whereas changes in the spectral index alter both the amplitude and shape of the oscillatory bump and the high-$k$ power–law tail. Because a blue‐tilted CDI component drives up small‐scale power very steeply, the allowed variation of the spectral index $n_{\rm cdi}$ around its fiducial value must be small for any given $(m_{\rm WDM},~f_{\rm WDM})$. Hence, each point in the WCDM $(m_{\rm WDM},~f_{\rm WDM})$ plane admits only a narrow band of CDI parameters that yield acceptable transfer‐function behavior, which generate small-scale power that is very similar to the adiabatic $\Lambda$CDM model.\footnote{
While one might imagine raising $n_{\rm cdi}$ above our fiducial value to further compensate WDM suppression, doing so produces an overshoot $T(k)\gg1$ and increasingly pronounced power on small scales. Current small-scale probes (e.g. Lyman-$\alpha$ flux power and UVLF) constrain both the magnitude and the shape of 3D power to a factor of few above $\Lambda$CDM. Taken together with the need for a minimum compensation, this yields a narrow allowed band of CDI spectral indices and amplitude.
} 

We conclude this section by noting that for representative values such as $f_{\rm cdi}(1 - f_{\rm wdm}) \sim \mathcal{O}(0.25)$, as discussed in the previous examples, the primordial isocurvature amplitude at the pivot scale $k_p = 0.05~\mathrm{Mpc}^{-1}$ is given by

$$
A_{\rm cdi}(k_p) \sim \mathcal{O}(0.06)\,(1 - f_{\rm wdm})^{-2}\,A_{\rm ad}(k_p).
$$

Assuming that the CDM consists predominantly of a single species, such as an axion-like particle (ALP), this amplitude can be extrapolated to smaller scales using the spectral tilt $n_{\rm cdi} > 1$, yielding

$$
A_{\rm cdi}(k) \approx \mathcal{O}(0.06)\,(1 - f_{\rm wdm})^{-2} \left( \frac{k}{k_p} \right)^{n_{\rm cdi} - 1} A_{\rm ad}(k_p).
$$

For our fiducial choice $n_{\rm cdi} \sim 3$ and $f_{\rm cdi}(1 - f_{\rm wdm}) \sim 0.25$, this implies that the CDI amplitude at $k = 100~h\,\mathrm{Mpc}^{-1}$ reaches
$$
A_{\rm cdi}(k=100~h\,\mathrm{Mpc}^{-1}) \sim \mathcal{O}(10^{-4})(1 - f_{\rm wdm})^{-2}.
$$
Consequently, for $f_{\rm wdm} = 0.99$, the primordial isocurvature amplitude can be of order unity, while for $f_{\rm wdm} = 0.9$, it remains as small as $\mathcal{O}(10^{-2})$.

However, we should note that from a viewpoint of models generating blue-tilted isocurvature fluctuations, there would be a cut-off in the power spectrum at which the scale dependence disappears. For example, the amplitude of a conserved, blue-tilted CDM isocurvature mode generated during inflation from an ALP field takes the approximate form
$$
A_{\rm cdi}(k) =
\begin{cases}
\left( \dfrac{k}{k_c} \right)^{n_{\rm cdi} - 1} \left( \dfrac{H}{\pi \theta f_a} \right)^2, & k < k_c \\
\left( \dfrac{H}{\pi \theta f_a} \right)^2, & k \geq k_c
\end{cases}
$$
where $H$ is the Hubble parameter during inflation, $f_a$ is the axion decay constant, $\theta$ is the initial misalignment angle, and $k_c$ corresponds to the comoving scale that exits the horizon at the time when the radial component of the ALP field relaxes to its vacuum \cite{Kasuya:2009up,Chung:2015tha}. For values of $H/f_a \gtrsim \mathcal{O}(10)$ to ensure that the $U(1)$ symmetry associated with the ALP is not restored by the de Sitter fluctuations, and for a tuned misalignment angle $\theta$, the inflationary isocurvature amplitude can naturally match the values required in our fiducial scenarios, particularly for regions with $f_{\rm wdm}\lesssim 0.9$, without violating perturbativity bounds on the linear isocurvature spectrum.

\section{WCDM with CDI: nonlinear analysis}\label{sec:WCDM_NL}
In the previous section, we showed that a partial compensation of the WDM free-streaming suppression with a blue-tilted CDI enhancement leaves small, oscillatory residuals in the linear mixed IC transfer function. One might worry that these linear residuals could lead to detectable features or suppression when projected into the full nonlinear analyses. To assess this, we now turn to the fully non-linear evolution of structure via $N$-body simulations.

We restrict ourselves here to DM-only simulations for two main reasons. First, the primary goal is to assess whether gravitational nonlinearities alone erase the $\mathcal{O}(0.1)$ linear residuals close to the WDM suppression scales as discussed in previous section and shown in Figs.~\ref{fig:TWCDMiso-2and3} and \ref{fig:TWCDMiso-4}, since baryonic physics (gas pressure, cooling, star formation, feedback) predominantly reshapes the matter distribution at smaller scales and introduces additional uncertainties that would obscure this pure-gravity test. Second, DM-only runs are computationally far less expensive, allowing us to explore a broader range of WDM masses, CDI tilts, and compensating amplitudes with high particle resolution.

We acknowledge a few limitations of this approach:
\begin{itemize}
    \item \textit{Neglect of baryons and thermal broadening}: real observations such as  Lyman-$\alpha$ forest statistics depend sensitively on gas temperature, pressure smoothing, and peculiar-velocity gradients.
    \item \textit{Astrophysical feedback}: Processes such as galactic winds or AGN feedback can further redistribute matter at intermediate scales.
    \item \textit{Halo/filament collapse history}: Even if the $P(k)$ is matched at lower redshifts, the timing of collapse can shift, which can feed into the IGM history. This could in turn alter the local temperature-density relation.
    \item \textit{UV background inhomogeneity}: If the CDI boosts small-scale clustering at early times, it could alter the spatial fluctuations in the UV background or patchiness of hydrogen reionization.
    \item \textit{1D flux power}: Converting 3D $P(k)$ into a 1D flux power spectrum involves velocity statistics and gas physics, thus requiring hydro-simulations or semi-analytic mapping.
\end{itemize}

Nonetheless, a DM-only demonstration that non-linear gravity alone washes out the residual oscillations is a powerful preliminary consistency check. It avoids the extra noise and computational cost of a full hydro-simulation, since gas physics would apply ``almost'' equally to CDM, WDM and WCDM+CDI once the DM scaffold is in place. In future work, we will extend this analysis to full hydrodynamics; here, however, we establish the crucial baseline that pure $N$-body dynamics already suppresses the residuals we highlighted in the linear theory.

\subsection{Simulation setup}\label{sec:sim_setup}
We perform DM-only $N$-body simulations using \textsf{FastPM} \cite{Feng:2016yqz} which is an approximated particle mesh $N$-body solver that implements the particle mesh (PM) scheme where the correct linear displacement evolution is enforced via modified kick and drift factors. 
The output phase space data from \textsf{FastPM} is processed using \textsf{nbodykit} \cite{Hand:2017pqn} to generate the nonlinear power spectra and halo mass function. We compute the power spectrum by deconvolving the effects of the interpolation on the measured power using \textit{compensated=True} for the TSC window, and reduce the contribution from aliasing (caused by finite mesh sampling) by keeping \textit{interlaced=True} option.

For each cosmological scenario such as a WCDM model with specified ICs, we employ three independent $N$-body simulations with box-size $L$ (in Mpc/h) and particle count per dimension $N_p$ given by ($L$,$N_p$) values $(80,1024)$, $(20,256)$, and $(20,1024)$. To meet the competing demands for a large box size and high resolution, we follow the approach of Refs.~\cite{McDonald:2001fe,Borde:2014xsa} and splice/combine the flux power spectrum from these three runs to construct a single nonlinear matter power spectrum spanning the wavenumber range $k\in [0.08,160]$ h/Mpc. The upper limit corresponds to the Nyquist frequency of the highest resolution box, with $(L,N_p)=(20,1024)$.

We generate the initial conditions for our $N$-body simulations at $z=29$ using second-order Lagrangian perturbation theory (2LPT) as implemented within \textsf{FastPM}, and evolve the system with 160 linear time-steps in the scale factor down to the redshift $z=3$. We do not include the thermal velocity of the WDM particles as it has been shown in Ref.~\cite{Leo:2017zff} that adding thermal velocity to the peculiar velocity of the particles adds more noise to the simulation which affects the matter and velocity power spectra at late times through forward propagation. We measure and present the power spectrum and halo mass function at redshifts $z=5.4,~4.2,\rm{and}~3$, which cover the broad redshift bins probed by current low- to high-resolution Lyman-$\alpha$ forest observations \cite{Viel:2013fqw}.

In CDM cosmology the linear growth function $D(k,z)$ remains scale-independent to better than one percent for wavenumbers $k\lesssim200\,h\,\mathrm{Mpc}^{-1}$; the tiny residual tilt at higher $k$ is generated only by the pressure of the sub-dominant radiation component. WDM particles, however, retain a finite thermal velocity dispersion, introducing a free-streaming (Jeans-like) pressure term in the linearised Euler equation.  Perturbations with $k\gtrsim k_{\mathrm{FS}}$ therefore grow more slowly than their CDM counterparts. For a DM-only $N$-body simulation that treats its particles as pressureless, the predicted power spectrum is reliable only on scales where the WDM growth function is still essentially CDM-like.  In this work the lightest species we consider has $m_{\mathrm{WDM}} = 0.6\;\mathrm{keV}$; for that mass and a warm-fraction $0.99$, the total growth lag compared to a CDM-cosmology remains below $5\,\%$ at $k\lesssim 100\,h\,\mathrm{Mpc}^{-1}$ between redshifts of $z=29$ to $z=0$. We therefore restrict our quantitative analysis to this wavenumber range.\footnote{Due to non-linear mode coupling, power on small scales with $k\gtrsim 100\,h\,\mathrm{Mpc}^{-1}$ receives substantial contributions from larger wavelengths, which mitigates the impact of the scale-dependent growth differences between CDM and WDM compared to the linear prediction.}

\subsection{Nonlinear results}\label{sec:NL_results}
In this section, we present nonlinear results from $N$-body simulations exploring the cosmological models discussed thus far. Below, we present a representative subset chosen for clarity and relevance of comparison. These models are listed in Table~\ref{tab:model_summary} and cover the warm dark matter fractions, particle masses, and initial condition types considered in fiducial examples presented in previous sections. In WCDM models, only the parameters listed in Table~\ref{tab:model_summary} are adjusted; the remaining parameters are kept as in the standard $\Lambda$CDM model.

\begin{table}[t]
\centering
\renewcommand{\arraystretch}{1.3}
\begin{tabular}{|p{5cm}|p{2cm}|p{2.5cm}|p{2.5cm}|p{3.5cm}|}
\hline
Model & $f_{\rm WDM}$ & $m_{\rm WDM}$ [keV] & ICs & $n_{\rm cdi},~f_{\rm cdi}(1 - f_{\rm WDM})$ \\
\hline
$\Lambda$CDM        & 0.00 & --   & Adiabatic & -- \\
WDM 6keV               & 1.00 & 6.0  & Adiabatic & -- \\
WCDM 1keV              & 0.14 & 1.0  & Adiabatic & -- \\
WCDM 600eV             & 0.08 & 0.6  & Adiabatic & -- \\
WCDM+CDI 1keV-I          & 0.99 & 1.0  & Mixed     & 2.83,\;0.293 \\
WCDM+CDI 1keV-II         & 0.99 & 1.0  & Mixed     & 2.92,\;0.220 \\
WCDM+CDI 600eV           & 0.99 & 0.6  & Mixed     & 3.00,\;0.313 \\
\hline
\end{tabular}
\caption{Summary of cosmological models. $f_{\rm WDM}$ is the warm dark matter fraction, $m_{\rm WDM}$ is the WDM particle mass, and ICs denotes the type of initial conditions. For mixed IC models, the spectral tilt $n_{\rm cdi}$ and effective isocurvature fraction $f_{\rm cdi}(1 - f_{\rm WDM})$ are also given.}
\label{tab:model_summary}
\end{table}

The adiabatic WCDM models are selected to reflect current lower bounds on the WDM particle mass as a function of the warm dark matter fraction, $f_{\rm WDM}$. The \textsf{WDM 6 keV} case represents a pure WDM model near the current lower mass limit. The \textsf{WCDM 1 keV} and \textsf{WCDM 600 eV} cases represent mixed models with warm fractions constrained to be $\lesssim 10\%$, where the particle masses are fixed to representative benchmark values. 

To highlight the effect of CDM isocurvature, we consider three WCDM models with mixed initial conditions and nearly pure WDM compositions ($f_{\rm WDM}=0.99$), corresponding to particle masses of $1$ keV and $600$ eV.  The \textsf{WCDM+CDI $1$keV-II} model has a slightly weaker CDI amplitude compared to the \textsf{$1$keV-I} model. These scenarios serve to underscore our main result: the inclusion of a blue-tilted CDI component can enable models with low-mass WDM and nearly full warm fractions to remain viable, thus effectively lifting the upper bound on $f_{\rm WDM}$ from $\sim 0.1$ to nearly unity, without violating current constraints on certain nonlinear observables. This provides a novel and theoretically motivated pathway to reconcile small-scale structure with otherwise excluded regions of WDM parameter space.

\subsubsection{Matter power spectrum}\label{sec:Tk_NL}
\begin{figure}[h]
    \centering
\includegraphics[width=1\linewidth]{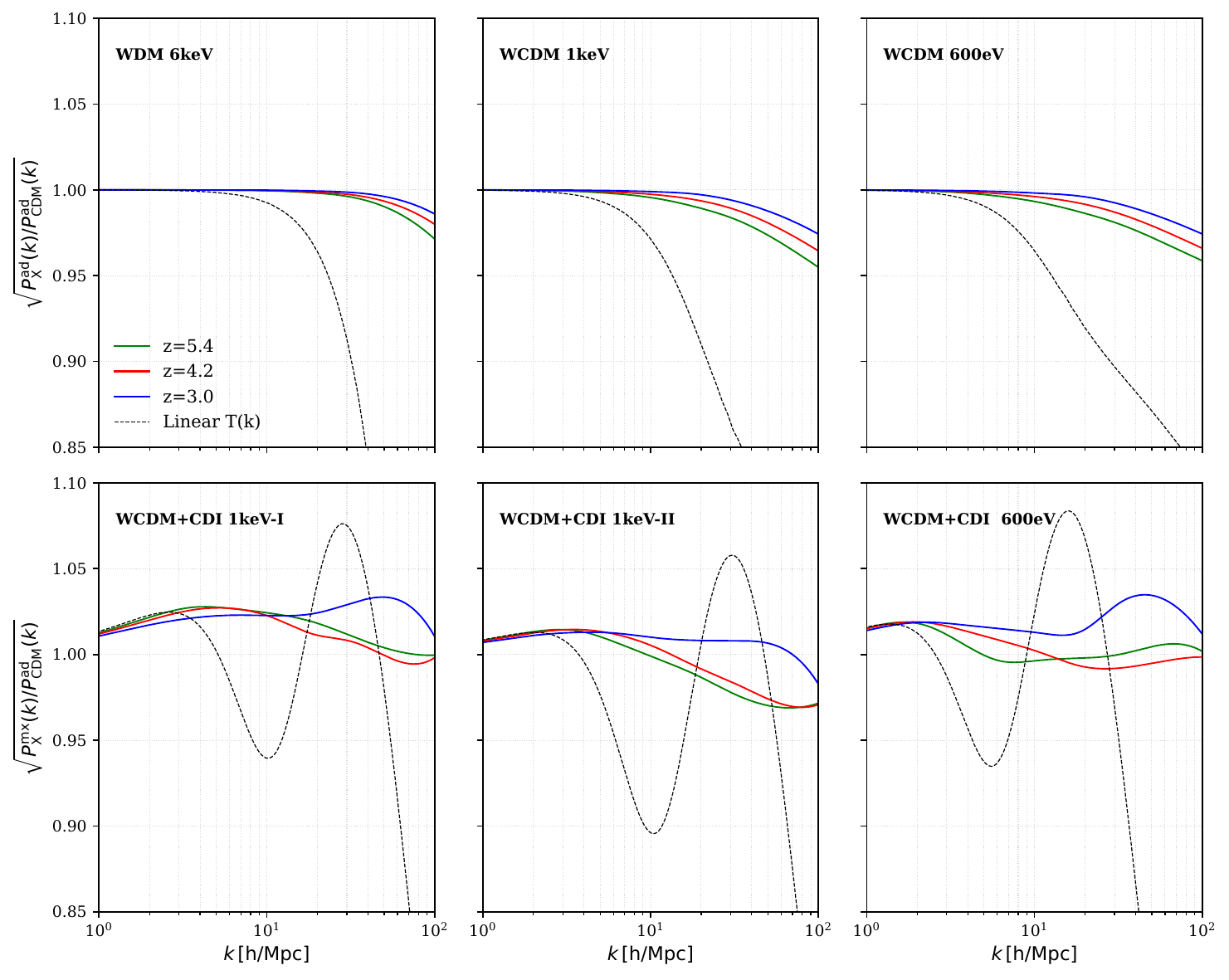} 
\caption{Matter transfer functions $T(k)$ from DM-only $N$-body simulations of the various non-$\Lambda$CDM cosmologies summarized in Tab.~\ref{tab:model_summary}. The solid green, red, and blue curves show the non-linear transfer function $T_{\rm nl}(k)$ at redshifts $z = 5.4$, $z = 4.2$, and $z = 3.0$, respectively. The dashed black curve represents the linear transfer function $T_{\rm lin}(k)$ evaluated at $z = 0$.}
    \label{fig:TNL}
\end{figure}

By evolving our WCDM models with initial conditions forward in time under gravity alone, we measure the nonlinear matter transfer function, defined as the ratio of the nonlinear matter power spectrum relative to $\Lambda$CDM.
We compare this across the set of WCDM cosmological models listed in Table~\ref{tab:model_summary}. Fig.~\ref{fig:TNL} presents resulting transfer functions at redshifts of $z=5.4$, $4.2$ and $3.0$, covering the range relevant for Lyman-$\alpha$ forest observations. In each panel, solid curves represent the smooth-interpolated nonlinear transfer functions, while dashed curves denote the corresponding linear transfer functions.

In all models, one sees the nonlinear gravitational clustering partially ``fills in'' the deficit, thereby damping (or smoothing out) the residual oscillatory features present in the linear transfer functions for the mixed IC WCDM models. For the three WCDM+CDI models considered here, the nonlinear transfer functions $T_{\rm nl}(k)$ sit close to unity (within only a few percent) for the entire redshift range $z\lesssim 5$. For the specific fiducial choice of CDI in the \textsf{$1$keV-II} model, the resulting nonlinear transfer functions closely track those of the WCDM models within the $k$ range $10-100 \,h\,\rm{Mpc}^{-1}$. Crucially, the curves highlight that the presence of a suitable isocurvature component in the initial conditions can modify the growth of structure in a way that partially compensates for the free-streaming damping characteristic of WDM, thus preserving more small-scale power, and shifting the characteristic
nonlinear cutoff scale to higher wavenumbers. Consequently, by adjusting the remaining cosmological parameters such as the adiabatic tilt $n_s$ and the relevant astrophysical nuisance parameters (for example, the IGM temperature-density relation parameters $T_0$ and $\gamma$ used when mapping the 3D nonlinear matter power spectrum into the 1D Lyman-$\alpha$ flux power), one can bring the WCDM+CDI predictions into close agreement with those of pure adiabatic WCDM or $\Lambda$CDM, with deviations of only a few percent, well within the current observational uncertainty of $\mathcal{O}(5-20)\%$. However, the distinct redshift-dependent growth and the subsequent smoothing of these oscillatory features can leave measurable signatures that may allow such a scenario to be ruled out.

At higher redshifts, the solid curves which represent the nonlinear transfer functions approximately track the shape of the linear ones (dashed curves) but are systematically shifted toward larger $k$.  Physically, this shift arises because mode-mode coupling drives power from large to small scales once structure formation goes nonlinear, so the characteristic suppression moves to smaller physical scales. As $z \rightarrow 0$, the transfer functions for the various IC models converge approximately. This convergence occurs since, post-virialization, the nonlinear matter distribution becomes increasingly insensitive to the detailed nature of the initial linear transfer function.

Finally, we note that gravitational nonlinearities tend to smooth out the detailed features of the linear transfer function. As such, cosmological observables at lower redshifts, especially those derived from the nonlinear matter distribution, become less sensitive to the presence of isocurvature modes. Conversely, high-redshift data analyses (e.g., UV luminosity functions \cite{Rudakovskyi:2021jyf,Sabti:2021unj}) that probe closer to the linear regime may be more effective at isolating the imprint of a CDI component in the initial conditions.

\subsubsection{Halo mass function}\label{sec:hmf}
\begin{figure}[h]
    \centering
    \includegraphics[width=1.\linewidth]{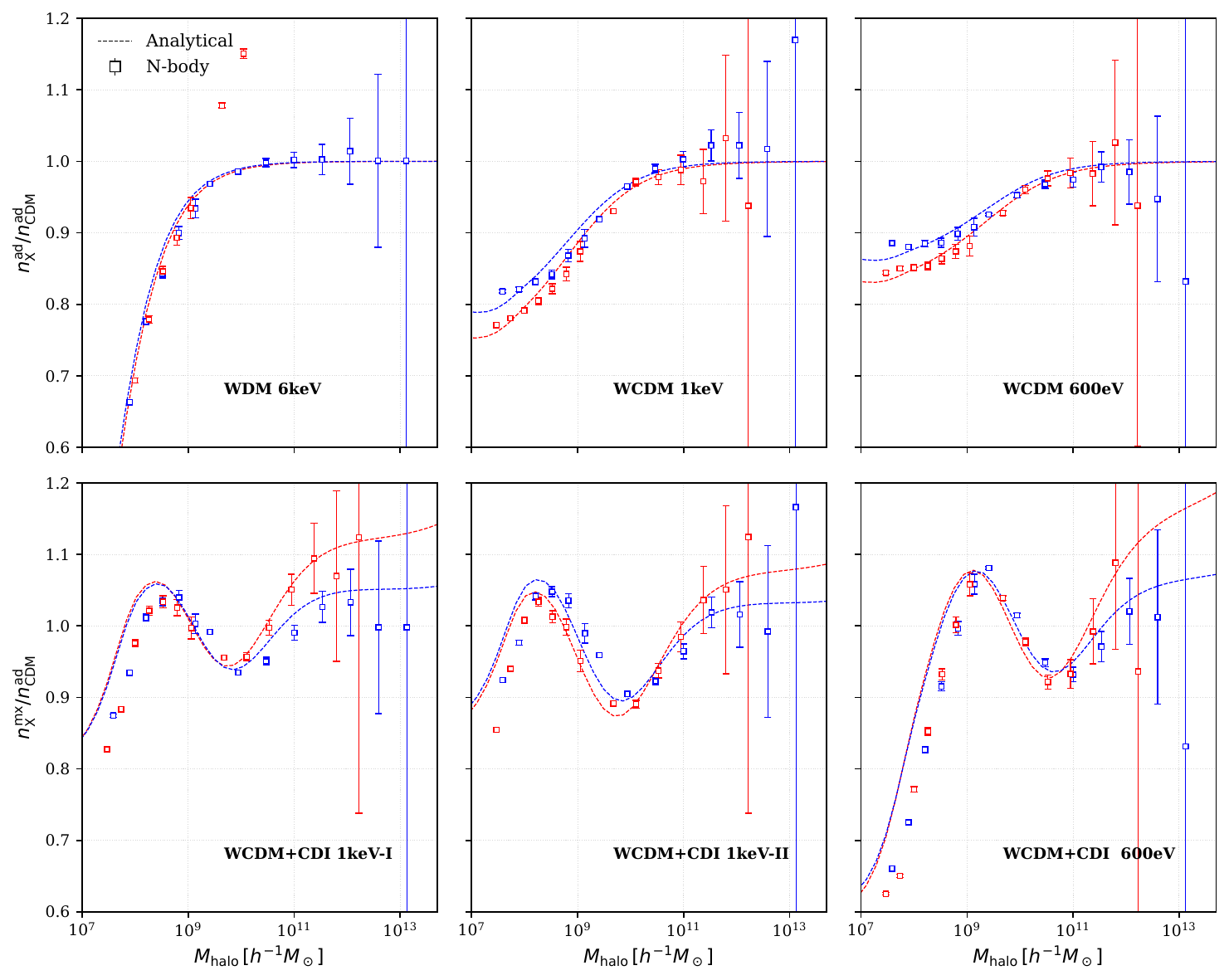}
    \caption{Halo mass functions for the non-$\Lambda$CDM scenarios with respect to the $\Lambda$CDM cosmology at redshifts of  $z=3.0$ (in blue) and $z=5.4$ (in red). The square markers denote the $N$-body result using FoF algorithm, while the dashed curves represent the theoretical predictions using the excursion-set theory. The WDM masses, warm fraction and the ICs for the various models are listed in Table~\ref{tab:model_summary}. In the WCDM+CDI models considered here, the mild $\sim1-2\%$ bump in the linear matter transfer functions at wavenumbers $0.5\lesssim k\,\rm{Mpc} \lesssim 3$ results in HMF ratios greater than $1$ on halo mass range above $10^{11}\,h^{-1}\,M_{\odot}$. We explain this effect in the main text.} 
    \label{fig:HMF}
\end{figure}

Finally, we compare the halo mass function (HMF) obtained from the $N$-body simulations for the various models highlighted in the previous subsection and listed in Table~\ref{tab:model_summary}. We also plot the approximate analytical HMFs obtained from the Sheth-Tormen ellipsoidal collapse parametrization as given in Ref.~\cite{Sheth:2001dp}. In this framework, the number of halos per unit mass per unit volume is expressed as
\begin{equation}
    \frac{dn}{d\ln M} = -\frac{\bar{\rho}}{M} f(\nu) \frac{d \ln \sigma}{d\ln M}\label{eq:HMF}
\end{equation}with
\[f(\nu) = A\sqrt{\frac{2q\nu^2}{\pi}}[1+(q\nu^2)^{-p}]e^{-q\nu^2/2}\] and where the linear mass variance 
\begin{equation}
    \sigma^2(M) = \int \frac{dk~k^2}{2\pi^2}P_{\rm lin}(k)W^2(kR).
\end{equation}
In Eq.~(\ref{eq:HMF}), $\nu=\delta_c/\sigma(M)$ where $\delta_c\approx 1.686$ is the linearly extrapolated spherical overdensity for
collapse.

The values of parameters $p,~A\equiv A(p),~\mbox{and }q$  are generally cosmology dependent. In Ref.~\cite{Gavas:2022iqb}, the authors showed that the HMF given in Eq.~(\ref{eq:HMF}) is not universal and has power spectrum dependence. For instance, the corresponding values for the standard $\Lambda$CDM cosmology respectively are $0.3, ~0.3222, ~\mbox{and }0.707$. The fitted parameters $p,~q$ show slight power-law spectral index dependence. Though for the WDM cosmology, the most critical change is through the modification of the window function $W(kR)$. HMF evaluations for standard $\Lambda$CDM models are generically performed using a top-hat window function. However, exercises in this direction (\cite{Schneider:2013ria,Schneider:2014rda,Benson:2012su,Leo:2018odn}) have shown that the top-hat filter is not suitable for WDM models with sharp power suppression. In this work, we consider a sharp-$k$ filter 
\[W(kR) = \Theta(1 - kR)\] from Ref.~\cite{Schneider:2013ria} for the \textsf{WDM 6 keV} model only (due to its sharp suppression in power) with parameters $p=0.3,~q=0.75$ to estimate analytical HMFs. For the rest of the models, we find that the top-hat filter with $q=0.707$ gives a better approximation. To achieve a well-defined mass within the sharp-$k$ filter, one should also modify the mass assignment to the halo radius $R$ as
\[ M(R) = \frac{4\pi}{3}\bar{\rho}(c R)^3\]where we take $c\approx 2.5$ in our work.

In Fig.~\ref{fig:HMF}, we plot the relative shape of the HMFs with respect to the $\Lambda$CDM case at redshifts $z=3.0$ (in blue color) and $z=5.4$ (in red color). The square markers represent the measurement from $N$-body simulations using the Friends-of-Friends (FOF) algorithm, with a linking length 0.2 times the mean interparticle separation. The dashed curves are obtained from the analytical HMF expression given in Eq.~(\ref{eq:HMF}) using a top-hat filter for all models, except the \textsf{WDM 6 keV} model for which we utilize a sharp-$k$ filter.

We find that the analytical HMF curves reproduce the nonlinear $N$-body HMF ratios to within a few percent across the halo mass range shown in Fig.~\ref{fig:HMF}, except for some spurious error in the HMF simulation data for \textsf{WDM 6keV} model at $z=5.4$. A noticeable discrepancy between analytical curves and simulation HMF points occurs for the WCDM+CDI models, whose residual oscillations in the linear transfer function lead the analytic prediction to overestimate abundances at the low‑mass end, specifically beyond the mass scale where the HMF ratio smoothly rises to a broad maximum and then decreases. By selecting an appropriate window function and fine‑tuning the fitting parameters $p$, $q$, and $c$, this overprediction can be alleviated. We address these calibrations in detail in a companion paper \cite{TSC}. 

Furthermore, in the mixed WDM+CDI models studied here, the HMF is about $10\%$ higher than in the fiducial $\Lambda$CDM case for halos with $M_{\rm halo} \gtrsim 10^{11}\,h^{-1}M_\odot$. 
The root cause is a modest $1–2 \%$ bump in the linear matter transfer function at wavenumbers $0.5\lesssim k/(\text{Mpc}^{-1})\lesssim3$ in our fiducial WDM+CDI scenarios (see Fig.~\ref{fig:TNL}). This feature slightly increases the variance of the density field on the corresponding mass scales. At early times, when the mean variance is lower overall, that small boost is amplified by the exponential sensitivity of halo abundance, yielding the $\mathcal{O}(10\%)$ surplus of massive haloes. 
This excess is most pronounced at early times, when massive halos are intrinsically rare and their number counts suffer from large sample-variance uncertainties \cite{Bouwens_2017}.
The complex structure of the HMF in our WCDM+CDI models may help address small-scale problems, such as those highlighted by recent JWST observations \cite{Castellano_2022,Boylan-Kolchin:2022kae,Carniani_2024}.

The mixed WCDM+CDI also exhibit oscillatory features in the HMF relative to $\Lambda$CDM, with amplitudes comparable to those seen in the linear matter transfer function. In all three mixed WCDM+CDI models, the HMF lies above that of the \textsf{WDM 6 keV} model in the low-mass $\lesssim 10^9\,h^{-1}M_\odot$ (subhalo) regime. Although the HMF is not directly observable, it underpins many key astrophysical observables that are sensitive to the abundance and distribution of dark matter halos. Observables such as the ultraviolet luminosity function of high‑redshift galaxies \cite{Rudakovskyi:2021jyf,Sabti:2021unj} and galaxy cluster counts \cite{Sartoris:2015aga} probe the low‑mass end of the halo mass function at $z\gtrsim3$, while strong‐lensing statistics \cite{Gilman:2019nap, Gilman:2021gkj} and cluster abundances constrain its high‑mass tail. Moreover, a detailed analysis of $N$‑body snapshots can extract the subhalo mass function and mass fractions, which in turn underpins Milky Way satellite galaxy counts \cite{Nadler:2021dft,Tan:2024cek}. Beyond mere counts, the \emph{internal kinematics} of these satellites (specifically the dynamical mass enclosed within their three-dimensional half-light radius), trace halo concentration enabling tests of the linear power spectrum on comoving scales of $k\sim 40\,\rm{Mpc}^{-1}$ \cite{Esteban:2023xpk} and extending to 
$k\sim100\,\rm{Mpc}^{-1}$ when the faintest ultra-faint dwarfs are included \cite{Dekker:2024nkb}. Deviations in the HMF due to WDM free-streaming and modifications from isocurvature modes can therefore leave measurable imprints in these observables. As such, the relative differences in the HMF presented here serve as an important theoretical input to forward-model the impact of non-standard dark matter physics on the galaxy population and their evolution. Nevertheless, they present a crucial theoretical signature that may become testable with the next generation of small-scale structure surveys. A careful mapping between the HMF and these observables is essential to translating our simulation-based results into robust empirical constraints on WCDM+CDI models.


\section{Discussion}\label{sec:conclusion}

In this study, we presented a minimal extension to the mixed warm and cold dark matter framework, highlighting how a blue-tilted, uncorrelated cold dark matter isocurvature component can effectively counterbalance the small-scale suppression typically associated with the WDM models. Through detailed analyses employing linear perturbation theory and $N$-body simulations, we demonstrated that even a $1\%$ CDM fraction with suitable isocurvature perturbations can significantly restore small-scale power suppression caused by free-streaming of WDM particles. Remarkably, this could permit notably lighter WDM particle masses, as low as 600 eV, to be compatible with stringent cosmological observations, including Lyman-$\alpha$ forest data and halo abundance constraints. Lighter WDM particles can give rise to cored density profiles, shaping the inner structure of halos, while the CDI component governs the form of the matter power spectrum on small scales, thereby enabling a possible decoupling of core formation from small-scale power suppression. While this study has focused on percent‑level CDM admixtures, larger CDM fractions with isocurvature could, in principle, enable substantially lower WDM particle masses.

Looking forward, several avenues merit further exploration. First, fully coupled hydrodynamic simulations (\cite{Paduroiu:2022uyb}) that self‐consistently include baryonic physics, gas cooling/heating, and feedback processes are essential to validate the WCDM+CDI model against observables of the intergalactic medium and galaxy formation. Extending these simulations to incorporate high-redshift galaxy luminosity functions, strong gravitational lensing statistics, and 21-cm tomographic data will tighten constraints on the mixed dark matter parameter space. Given the high dimensionality introduced by additional isocurvature parameters, one practical strategy is to construct a model-independent transfer-function emulator by running $\mathcal{O}(10^2)$ representative WCDM+CDI cosmologies, train an emulator, and employ it within an MCMC framework for rigorous Bayesian inference using current and forthcoming survey data.

A more immediate and computationally economical analysis can leverage analytic halo mass functions based on linear transfer functions to model halo-based observables, such as existing UV luminosity function likelihoods, and thereby derive preliminary bounds on the WDM particle mass and the warm fraction $f_{\rm WDM}$ in the presence of isocurvature perturbations. We are actively pursuing this approach and will present the results in an upcoming companion paper \cite{TSC}.

From a theoretical standpoint, the WCDM+CDI scenario also stimulates investigations into the physics of the early universe. In particular, identifying concrete mechanisms that generate strongly blue-tilted CDM isocurvature spectra and their attendant non‑Gaussian signatures would provide crucial model‐building guidance. Within the observationally allowed parameter regime, one could then devise explicit particle-physics realizations of mixed WCDM+CDI frameworks. A particularly interesting direction would be to investigate early-universe particle-physics models capable of constraining the amplitude and spectral index of the CDM isocurvature component, thereby reducing the effective freedom in the CDI parameter space.

Despite its promise, this hybrid model is not without caveats. While the warm component alleviates inner-halo density tensions via reduced concentrations and sizeable phase-space-driven cores, the residual cold fraction possesses effectively infinite primordial phase-space density. This can lead to central mass segregation and the re-emergence of cuspy inner profiles, potentially reinstating the TBTF and cusp-core challenges on subgalactic scales. Moreover, the CDM subcomponent with a large isocurvature power may form ultra-dense subhalos, whose signatures could manifest in strong-lensing flux‐ratio anomalies \cite{Kochanek:2003zc}, or even millilensing. Thorough validation will therefore require two-fluid simulations and a coordinated analysis of dwarf-galaxy kinematics, inner profile, lensing substructure, and high-redshift structure formation.

Ultimately, our work provides both observational cosmologists and theoretical physicists with an interesting framework that encourages deeper exploration of the interplay between primordial physics and small-scale structure formation. In particular, it utilizes the combined effects of primordial fluctuations and background cosmology in jointly shaping the evolution of structure on small scales. Through a combination of theoretical modeling, detailed simulations, and observational efforts, the WCDM+CDI model may provide a solution to persistent cosmological tensions, potentially deepening our understanding of the dark matter sector and the physics of the early universe.

\section*{Acknowledgement}
This research was supported in part by Lilly Endowment, Inc., through its support for the Indiana University Pervasive Technology Institute. T.T is supported by  JSPS KAKENHI Grant Numbers 25K01004 and MEXT KAKENHI 23H04515, 25H01543. T.S.C. thanks Keir K. Rogers for discussions on the compressed Lyman-$\alpha$ likelihood.

\nocite{apsrev41Control}
\bibliographystyle{JHEP2.bst}
\bibliography{refs}

\end{document}